\documentclass{article}

\usepackage{arxiv}

\usepackage[utf8]{inputenc} % allow utf-8 input
\usepackage[T1]{fontenc}    % use 8-bit T1 fonts
\usepackage{hyperref}       % hyperlinks
\usepackage{url}            % simple URL typesetting
\usepackage{booktabs}       % professional-quality tables
\usepackage{amsfonts}       % blackboard math symbols
\usepackage{nicefrac}       % compact symbols for 1/2, etc.
\usepackage{microtype}      % microtypography
\usepackage{graphicx}
\usepackage{natbib}
\usepackage{doi}

\usepackage{framed,multirow}
\usepackage{amssymb}
\usepackage{latexsym}
\usepackage{xcolor}
\usepackage{amsmath}
\graphicspath{ {./figures/} }
\usepackage{subcaption}
\usepackage{adjustbox}
\usepackage{longtable}
\usepackage{tabularx} % Allows for tabular cells to take X width
\usepackage{lineno}

\title{ATOMMIC: An Advanced Toolbox for Multitask Medical Imaging Consistency to facilitate Artificial Intelligence applications from acquisition to analysis in Magnetic Resonance Imaging}

\usepackage{authblk}

\author[1,2,3]{Dimitrios ~Karkalousos}
\author[1,2,4]{Ivana ~Išgum}
\author[1,2,3]{Henk A.~Marquering}
\author[1,3]{Matthan W.A.~Caan}
\affil[1]{Department of Biomedical Engineering \& Physics, Amsterdam University Medical Center, Location University of Amsterdam, Amsterdam, The Netherlands}
\affil[2]{Department of Radiology \& Nuclear Medicine, Amsterdam University Medical Center, Location University of Amsterdam, Amsterdam, The Netherlands}
\affil[3]{Amsterdam Neuroscience, Brain Imaging, Amsterdam, The Netherlands}
\affil[4]{Informatics Institute, University of Amsterdam, Amsterdam, The Netherlands}

\renewcommand{\shorttitle}{ }

\hypersetup{
pdftitle={ATOMMIC: An Advanced Toolbox for Multitask Medical Imaging Consistency to facilitate Artificial Intelligence applications from acquisition to analysis in Magnetic Resonance Imaging},
pdfauthor={Dimitrios ~Karkalousos, Ivana ~Išgum, Henk A.~Marquering, Matthan W.A.~Caan},
pdfkeywords={Deep Learning, Magnetic Resonance Imaging, Image reconstruction, Image segmentation, Quantitative MRI, Multitask Learning},
}

\date{}

\begin{document} % Add this line
\maketitle

\begin{abstract}
\textbf{Background and Objectives:} Artificial intelligence (AI) is revolutionizing Magnetic Resonance Imaging (MRI) along the acquisition and processing chain. Advanced AI frameworks have been developed to apply AI in various successive tasks, such as image reconstruction, quantitative parameter map estimation, and image segmentation. However, existing frameworks are often designed to perform tasks independently of each other or are focused on specific models or single datasets, limiting generalization. This work introduces the Advanced Toolbox for Multitask Medical Imaging Consistency (ATOMMIC), an open-source toolbox that streamlines AI applications for accelerated MRI reconstruction and analysis. ATOMMIC implements several tasks using deep learning (DL) networks and enables MultiTask Learning (MTL) to perform related tasks in an integrated manner, targeting generalization in the MRI domain. \textbf{Methods:} We first review the current state of AI frameworks for MRI through a comprehensive search of the existing literature and by parsing 12,479 GitHub repositories. We benchmark twenty-five deep learning (DL) models on eight publicly available datasets to present distinct applications of ATOMMIC on the tasks of accelerated MRI reconstruction, image segmentation, quantitative parameter map estimation, and joint accelerated MRI reconstruction and image segmentation utilizing MTL. \textbf{Results:} Our findings demonstrate that ATOMMIC is the only framework supporting MTL for multiple MRI tasks with harmonized complex-valued and real-valued data support. Evaluations on single tasks show that physics-based models, which enforce data consistency by leveraging the physical properties of MRI, outperform other models in reconstructing highly accelerated acquisitions. Physics-based models that produce high reconstruction quality can accurately estimate quantitative parameter maps. When high-performing reconstruction models are combined with robust segmentation networks utilizing MTL, performance is improved in both tasks. \textbf{Conclusions:} ATOMMIC facilitates MRI reconstruction and analysis by standardizing workflows, enhancing data interoperability, integrating unique features like MTL, and effectively benchmarking DL models. With ATOMMIC, we aim to provide researchers with a comprehensive AI framework for MR Imaging that can also serve as a platform for new AI applications in medical imaging.
\end{abstract}

% keywords can be removed

\section{Introduction}

In recent years, Artificial Intelligence (AI) has led to significant advancements in medical imaging, spanning various tasks along the acquisition and processing chain. Deep Learning (DL) segmentation Convolutional Neural Networks (CNNs) enable fast and accurate segmentation of anatomy and pathology in Magnetic Resonance Imaging (MRI) and Computed Tomography (CT) (\cite{isenseeNnUNetSelfconfiguringMethod2021, milletariVNetFullyConvolutional2016, oktayAttentionUNetLearning2018, ronnebergerUNetConvolutionalNetworks2015}). DL models have also been developed to act as inverse problem solvers, improving the reconstruction quality in MRI and CT (\cite{adlerLearnedPrimalDualReconstruction2018}). In MRI specifically, scanning time can be reduced considerably by accelerating the acquisition process, while DL reconstruction models offer high-quality images by training on the sparse signal representation (\cite{aggarwalMoDLModelBased2019, hammernikLearningVariationalNetwork2018, lonningRecurrentInferenceMachines2019}). Similarly DL models can accurately estimate quantitative parameter maps from multiple accelerated MRI acquisitions while varying sequence parameters (\cite{zhangUnifiedModelReconstruction2022}).

MultiTask Learning (MTL) can further improve the performance of models performing individual but related tasks by combining and jointly performing them (\cite{caruanaMultitaskLearning1997}). For instance, the segmentation efficacy depends on the reconstruction quality, as the former task invariably follows the latter. Therefore, since tasks are related, if they are performed simultaneously, performance can be improved on both while reducing the overhead time of performing the tasks separately. Although MTL has been successfully applied to combine reconstruction and segmentation (\cite{huangBrainSegmentationKSpace2019, karkalousosMultiTaskLearningAcceleratedMRI2024, pramanikRECONSTRUCTIONSEGMENTATIONPARALLEL2021, sunJointCSMRIReconstruction2019}), the challenge lies in maintaining consistency in performance while merging multiple single-task DL models and harmonizing data support for both complex-valued and real-valued domains. Dedicated medical imaging AI frameworks are usually employed to address the issue of regularization across tasks and data types. Primarily, such frameworks are task-specific, focusing on essential tasks like reconstruction (\cite{blumenthalDeepDeepLearning2023, yiasemisDIRECTDeepImage2022}) and segmentation (\cite{gibsonNiftyNetDeeplearningPlatform2018, isenseeNnUNetSelfconfiguringMethod2021, wangPyMICDeepLearning2023}), or they are modality-specific (\cite{thibeau-sutreClinicaDLOpensourceDeep2022, tournierMRtrix3FastFlexible2019, tustisonANTsXEcosystemQuantitative2021}), or focus on data pre-processing and data augmentations (\cite{perez-garciaTorchIOPythonLibrary2021}). The Medical Open Network for Artificial Intelligence (MONAI) (\cite{cardosoMONAIOpensourceFramework2022}) is a popular framework that supports multiple tasks, modalities, and data types. However, tasks can only be performed independently, and complex-valued data support is limited to the reconstruction task. 

Nevertheless, integrating multiple data types support or multiple methods for MTL in existing frameworks can be complicated due to differences in data structures, formats, and programming languages, increasing the burden for researchers who need a more comprehensive range of options for medical image analysis. To address these inconsistencies, we present the Advanced Toolbox for Multitask Medical Imaging Consistency (ATOMMIC), an open-source toolbox supporting multiple independent MRI tasks, such as reconstruction, segmentation, and quantitative parameter map estimation, and uniquely integrates MTL. Complex-valued and real-valued data support is unified, ensuring consistency among tasks, DL models, supported datasets, and training and testing schemes. In ATOMMIC, each task is implemented as a collection of data loaders, models, loss functions, evaluation metrics, and pre-processing transformations specific to each independent task. By using collections, new datasets, models, and tasks can be easily integrated. The MTL collection inherits the independent tasks and further extends them by combining related or subsequent ones, such as reconstruction and segmentation or reconstruction and quantitative parameter map estimation. ATOMMIC is built according to NVIDIA's NeMO (\cite{kuchaievNeMoToolkitBuilding2019}), a computationally efficient conversational AI toolkit that allows for high-performance training and testing using multiple GPUs, multiple nodes, and mixed precision support.

In the following sections, we first explore the landscape of AI frameworks for medical imaging (Sec. \ref{sec:related_work}) through a thorough literature search and, additionally, parsing GitHub repositories, showcasing the need for multitasking toolboxes that support multiple tasks and data types with detailed documentation and up-to-date maintenance. Next, we introduce ATOMMIC's main components (Sec. \ref{sec:methods}), including the supported tasks, pre-processing transforms, training options, DL models, and datasets. We benchmark twenty-five DL models on eight publicly available datasets, including brain and knee anatomies, for accelerated MRI reconstruction, quantitative MRI parameter map estimation, segmentation, and MTL. Various undersampling schemes and acceleration factors are tested in the reconstruction, quantitative parameter map estimation, and MTL tasks. At the same time, accurate segmentation of brain lesions, tumors, and knee pathologies is assessed in the segmentation and MTL tasks to demonstrate applications of ATOMMIC across different settings (Sec. \ref{sec:results}). Finally, in Sec. \ref{sec:discussion}, we discuss how ATOMMIC aims to provide a multitask toolbox for the research community to use, develop, and share models and potentially datasets and pre-processing pipelines across various MRI tasks, targeting generalization in the MRI domain. 

The datasets used in the experiments (Sec. \ref{sec:experiments}) are publicly accessible, while pre-processing pipelines, detailed API documentation, tutorials, and quick start guides are available on the open-source ATOMMIC repository\footnote{\url{https://github.com/wdika/atommic}}, under Apache 2.0 license. Trained models' checkpoints are available on HuggingFace\footnote{\url{https://huggingface.co/wdika}}, allowing for full reproducibility of the results. 

\section{Related work}
\label{sec:related_work}

\begin{figure}[!t]
    \centering
    \includegraphics[width=0.475\textwidth]{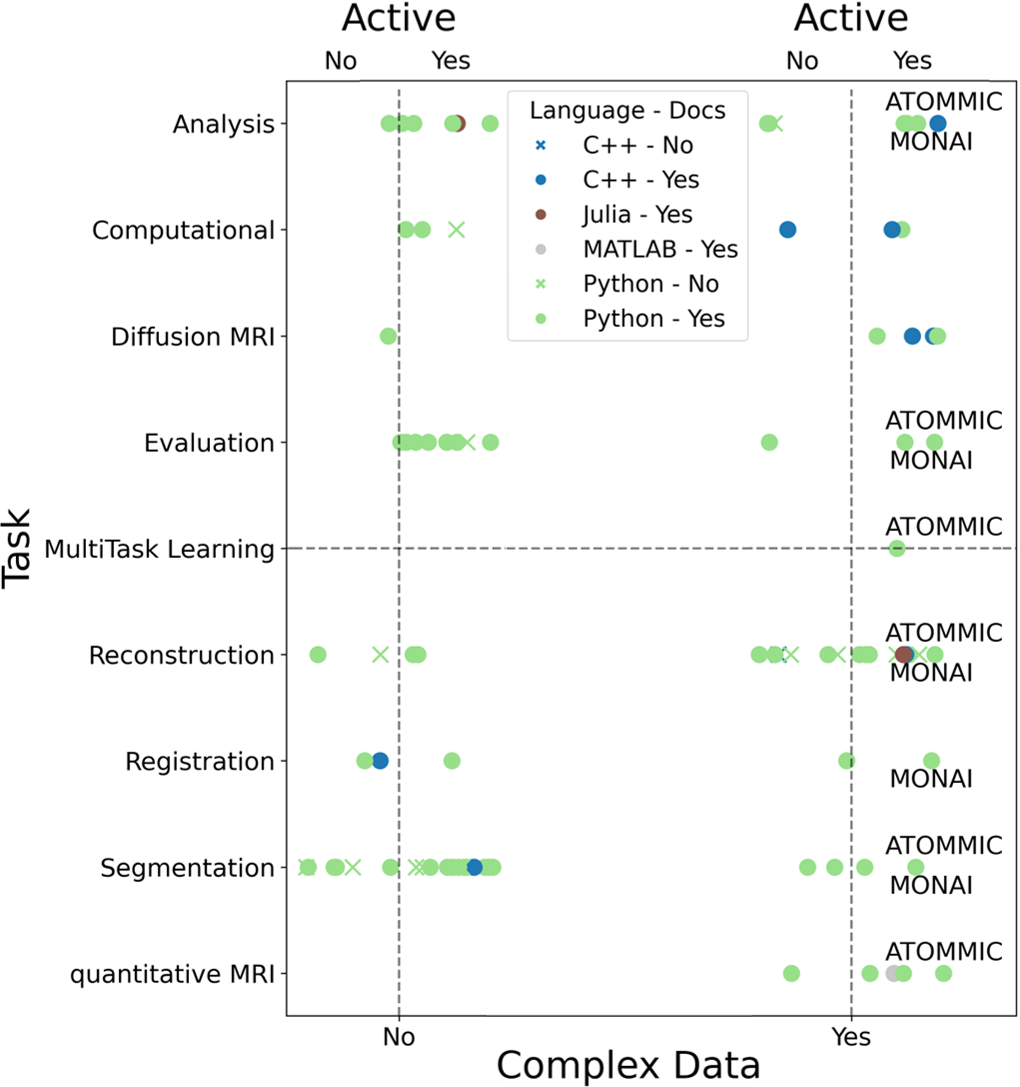}
    \caption{Overview of AI repositories for MRI tasks parsed from GitHub. The repositories are divided into two groups on the x-axis: those that do not support complex-valued data (left group) and those that do support complex-valued data (right group). The repositories are further categorized based on their activity level, on the top of the x-axis, split by vertical dashed lines: those that have committed updates within the last year (2023) are labeled as "Active Yes" (right side), and those that have not as "Active No" (left side). The supported tasks are depicted on the y-axis, while the horizontal dashed line showcases the multitasking toolboxes. Repositories are visualized as dots if documentation is available (Docs Yes) and cross marks if documentation is unavailable (Docs No). Each color signifies the language of the repository, with blue representing C++, brown representing Julia, gray representing MATLAB, and green representing Python.}
    \label{fig:repositories_overview}
\end{figure}

Recognizing that the emphasis in research is often placed on the model implementation or dataset specifics rather than on frameworks, we extended our literature search of AI frameworks for MRI to include GitHub repositories, not limited to those published in scientific papers. Utilizing keywords such as 'medical-image-processing', 'medical-imaging', 'MRI', 'medical', 'MRI-reconstruction', 'MRI-segmentation', 'neuroimaging', 'nifti', 'dicom', 'compressed-sensing', 'image-reconstruction', 'brain', 'medical-image-analysis', and 'MRI-registration', we identified a total of 12,479 repositories. Removing duplicates and non-existent URLs resulted in 10,747 repositories. Next, we defined a minimum usage threshold based on the number of stars, where a star serves as a popularity and usage metric on GitHub. The minimum number of stars was 2, the maximum was 23,400, and the median was 13. Limiting our results to repositories with at least ten stars returned 3,623 repositories. We meticulously narrowed this list to 68 DL frameworks pertinent to MRI. In brief, we removed repositories irrelevant to MRI, not written in English, and containing data and file converters only. Furthermore, we discarded Graphic User Interfaces, specific model and paper implementations, theses, lab pages, and courses. A detailed list of the repositories, including URLs, is available on GitHub\footnote{\url{https://github.com/wdika/atommic}}. This comprehensive review highlights a gap in frameworks supporting MultiTask Learning for MRI, complex-valued data support, providing documentation, and up-to-date maintenance, as shown in Fig. \ref{fig:repositories_overview}.

Among the 68 AI frameworks for MRI identified, ATOMMIC and MONAI were notable for their up-to-date maintenance, detailed documentation, and support for multiple independent tasks. However, as shown in Fig. \ref{fig:repositories_overview}, ATOMMIC emerged as the only toolbox supporting MTL with harmonized complex-valued and real-valued data support, comprehensive documentation, and up-to-date maintenance. 

\section{Methods}
\label{sec:methods}

\begin{figure*}[!t]
    \centering
    \includegraphics[width=1.0\textwidth]{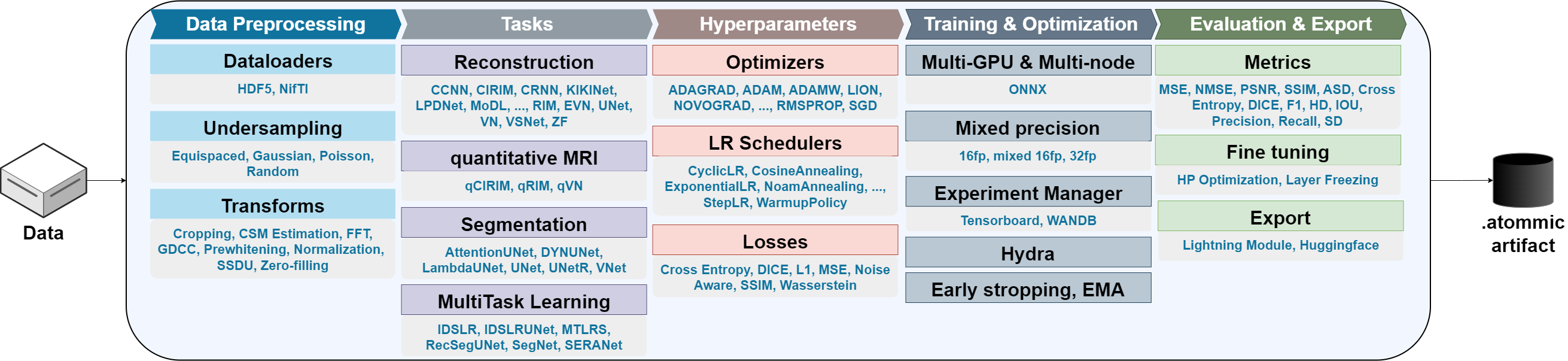}
    \caption{Schematic overview of ATOMMIC. Starting from the left to the right, MRI data are given as input. Next, configurations such as dataloaders, undersampling schemes, transforms, task(s) and models, optimizers, learning rate schedulers, losses, training and optimization settings, evaluation metrics, and exports are defined. The output is an atommic artifact (rightmost) containing the trained model's checkpoints and configurations, which can be directly used for inference on new datasets.}
    \label{fig:atommic_schematic_overview}
\end{figure*}

This section presents an overview of the MRI tasks supported in ATOMMIC, including accelerated MRI reconstruction, segmentation, quantitative parameter map estimation, and MTL for joint reconstruction and segmentation. Furthermore, we describe the process of training and testing DL models in ATOMMIC and showcase the available pre-processing transformations. A schematic overview summarizing ATOMMIC's features and workflow is included for enhanced comprehension (Fig. \ref{fig:atommic_schematic_overview}). Finally, we present benchmarks and use cases to showcase the toolbox's advantages and capabilities.

\subsection{MRI tasks}
\label{sec:mri-tasks}

Starting from the acquisition process, the forward model of acquiring and accelerating MRI data can be expressed as follows:
\begin{equation}
y_{i} = P \cdot \mathcal{F}\left(S_{i}^{H} \odot x_{i}\right) + \sigma_{i}, i=1,...,N,
\label{eq:mri-acq}
\end{equation}
where $x_{i}$ represents fully sampled multicoil complex-valued data for $N$ total coils. $S_{i}$ denotes the coil sensitivity maps, which homogenize the spatial intensities, $H$ is the Hermitian complex conjugate, and $\odot$ is the Hadamard product. $\mathcal{F}$ is the Fourier transform, projecting the data onto the frequency domain, known as k-space in MRI. $P$ is the undersampling scheme, which accelerates imaging by reducing the amount of data needed to acquire, and $\sigma_{i}$ represents the noise inherent in the acquisition process. The resulting undersampled multicoil data are denoted by $y_{i}$.

While the forward process is well-defined, reconstructing high-quality images from undersampled data or finding the inverse mapping process $y \mapsto x$ remains challenging. The inverse problem of accelerated MRI reconstruction can be solved through a Bayesian estimation. The goal is to maximize the posterior distribution of $y$ given $x$ and the prior probability of $x$. This process is known as the Maximum A Posteriori (MAP) estimation and can be described as:
\begin{equation}
x_{\text{MAP}} = \arg\max_x \left( \log p\left(y|x\right) + \log p\left(x\right) \right).
\label{eq:map}
\end{equation}
Substituting Eq. \ref{eq:mri-acq} into Eq. \ref{eq:map} transforms the inverse problem into a minimization problem:
\begin{equation}
x = \arg\min_{x} \left\{ \sum_{i=1}^{N} \theta\left(y_i, P \cdot F\left(S_{i}^{H} \odot x_i\right) + \sigma_i\right) + \lambda R\left(x\right) \right\},
\label{eq:optim}
\end{equation}
where $\theta$ represents the discrepancy between the undersampled measurements and their predictions. The summation indicates that the multicoil data are transformed into a coil-combined image. $\lambda$ denotes the weighting factor for the regularizer $R$. The regularizer can be modeled using a neural network.

Following accelerated MRI reconstruction by solving the inverse problem in Eq. \ref{eq:optim}, the task of estimating quantitative parameter maps can be expressed when data from multiple acquisitions with varying sequence parameters are available. The Multiple Echo Recombined Gradient Echo (ME-GRE) sequence varies the echo time (TE), such that the apparent transverse relaxation rate ($R^*_{2}$) may be computed from repeated acquisitions. The forward relaxation model describes the acquisition process for multiple TEs as
\begin{equation}
{x}_t = M \odot e^{-\mathrm{TE}_t \left(R^*_{2} - B_0i\right)},
\label{eq:qmri-acq}
\end{equation}
where $M$ is the net magnetization, $t$ denotes a single echo time, and $B_0$ is the off-resonance of the static magnetic field, and $i$ notates complex-valued data. Inserting the forward relaxation model (Eq. \ref{eq:qmri-acq}) into the forward model of accelerated MRI acquisition (Eq. \ref{eq:mri-acq}) results in a unified quantitative MRI forward model 
\begin{equation}
y_{t, i} = P \cdot \mathcal{F} \left(S_{i}^{H} \odot \left(M \odot e^{-\mathrm{TE}_t \left(R^*_{2} - B_0\right)}\right)\right) + \sigma_{i}.
\label{eq:qmri-forward-model}
\end{equation}
The resulting parameter maps follow from minimizing Eq. \ref{eq:optim}.

In MultiTask Learning (MTL), multiple related tasks are combined and performed simultaneously instead of individually, aiming to identify relationships, leading to better generalization and enhancing the performance of each task (\cite{adlerTaskAdaptedReconstruction2022}). For example, reconstruction can be combined with subsequent tasks, such as quantitative parameter map estimation or segmentation, by modeling the regularizer $R$ in Eq. \ref{eq:optim} with a neural network and approximating $x$ through an iterative training scheme. The predicted reconstruction, $\hat{x}$, is given as input to the subsequent task-specific network during each iteration. In the case of MTL for reconstruction and quantitative parameter map estimation, $\hat{x}$ is inserted into Eq. \ref{eq:qmri-forward-model}. When combining reconstruction with segmentation, $\hat{x}$ is mapped onto delineated anatomical structures. Features are shared, effectively acting as inductive bias for all tasks using a joint loss function (\cite{huangBrainSegmentationKSpace2019, karkalousosMultiTaskLearningAcceleratedMRI2024, pramanikRECONSTRUCTIONSEGMENTATIONPARALLEL2021, sunJointCSMRIReconstruction2019}).

\subsection{Undersampling MRI}
\label{sec:accelerated_mri}

Data undersampling, as described in Eq. \ref{eq:mri-acq} by the undersampling mask $P$, is crucial to accelerate the acquisition process by partially sampling or sub-sampling the k-space. Prospective undersampling refers to accelerating imaging during the data acquisition phase. Retrospective undersampling refers to generating undersampling masks post-acquisition and applying them to fully sampled data, usually for research purposes. ATOMMIC supports both prospective and retrospective undersampling. Each undersampling scheme is implemented in a respective class as follows.

For equispaced 1D undersampling, the \texttt{Equispaced1DMaskFunc} class is utilized to generate a mask with evenly spaced lines in the fully sampled k-space (\cite{muckleyResults2020FastMRI2021}). A number of fully sampled low frequencies in the center of k-space is defined as $N_{\text{low\_freqs}} = (N \cdot \texttt{center\_fractions})$, where $N$ is the size of the k-space and \texttt{center\_fractions} is a parameter that can be adjusted. The chosen \texttt{accelerations} define the resulting undersampling rate, equal to $\frac{N}{\texttt{accelerations}}$. For 2D equispaced undersampling, the \texttt{Equispaced2DMaskFunc} class provides similar functionality. For a more randomized approach, the \texttt{Random1DMaskFunc} class allows for 1D undersampling with random spacing of the sampled k-space lines.

In the case of Gaussian density weighted undersampling, the \texttt{Gaussian1DMaskFunc} class generates a Gaussian 1D mask. Data points are sampled based on the probability density function of the Gaussian distribution. The half-axes of the ellipse are set to the \texttt{center\_scale} percentage of the fully sampled region. The peripheral points are randomly sampled according to a Gaussian probability density function. Here, the \texttt{center\_fractions} equivalent is the Full-Width at Half-Maximum (FWHM). Similarly, the \texttt{Gaussian2DMaskFunc} class allows Gaussian 2D undersampling, where data points near the center of the k-space are fully sampled within an ellipse. The half-axes of the ellipse are set to the \texttt{center\_scale} percentage of the fully sampled region. The \texttt{Poisson2DMaskFunc} class allows for non-random sampling, generating a 2D mask following a Variable-Density Poisson-disc sampling pattern.

\begin{figure}[!t]
    \centering
    \includegraphics[width=0.98\linewidth]{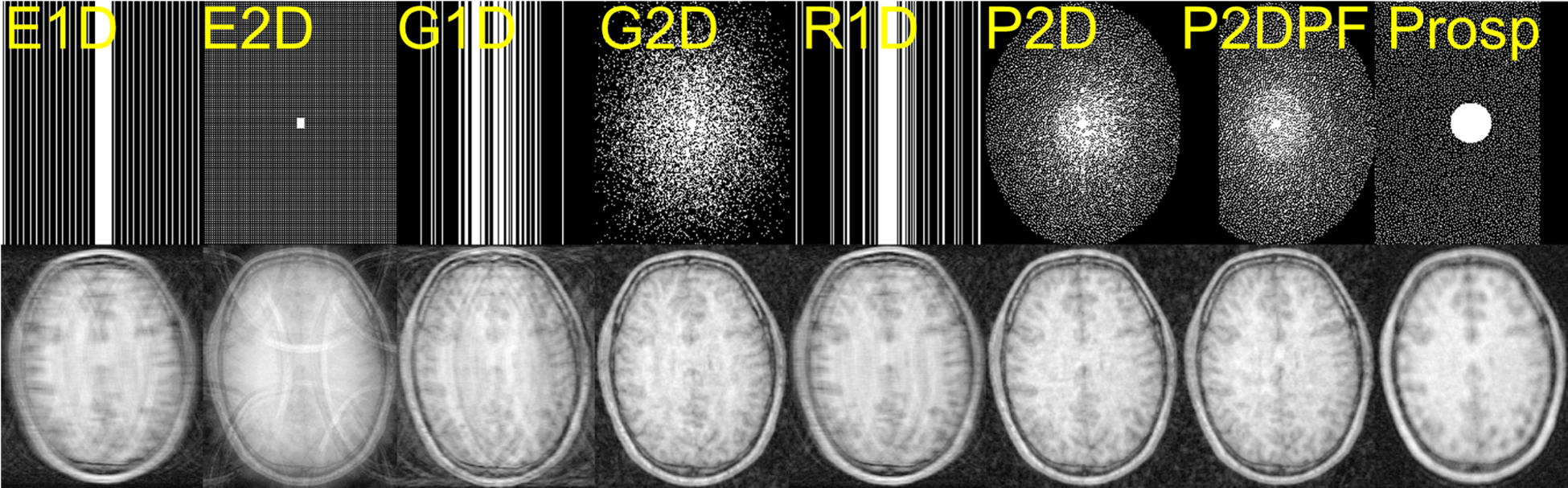}
    \caption{Overview of undersampling options in ATOMMIC. From left to right, columns one to seven present retrospective undersampling using Equispaced 1D (E1D), Equispaced 2D (E2D), Gaussian 1D (G1D), Gaussian 2D (G2D), Random 1D (R1D), Poisson 2D (P2D), and Poisson 2D with 20\% Partial Fourier (P2DPF) masking, respectively. Note that Partial Fourier can be applied to any masking. The last column presents prospective undersampling (Prosp) using the Calgary-Campinas 359 dataset default 2D Poisson mask (\cite{beauferrisMultiCoilMRIReconstruction2022}).}
    \label{fig:undersampling}
\end{figure}

The \texttt{partial\_fourier} parameter sets a percentage of outer k-space that is not sampled, resulting in a partially sampled k-space. An illustrative overview of the undersampling options is provided in Fig. \ref{fig:undersampling}.

\subsection{MRI transforms}
\label{sec:mri-transforms}

MRI transforms in ATOMMIC refer to pre-processing, i.e., data augmentations, intensity normalization, and multicoil-related transforms. The following transforms are implemented to handle both complex-valued and real-valued data for any task (Fig. \ref{fig:atommic_schematic_overview}).

The \texttt{NoisePreWhitening} class ensures that the inherent noise in the acquisition process, represented by $\sigma$ in Eq. \ref{eq:mri-acq}, will be independent and identically distributed by applying noise pre-whitening and decoupling or decorrelating coil signals \cite{hansenGadgetronOpenSource2013}. While MRI acquisitions commonly include separate noise measurements, such information is only sometimes exported. When this information is unavailable, the physical properties are modeled, assuming that the periphery of k-space is dominated by noise, such that a patch can be defined to measure the noise level. Its size can be set manually with the \texttt{prewhitening\_patch\_start} and \texttt{prewhitening\_patch\_length} parameters or automatically by toggling the \texttt{find\_patch\_size} parameter. Alternatively, if the actual noise level is already measured and is available, it can be given as input instead of measuring the noise level with a patch. Also, the \texttt{scale\_factor} parameter is used for setting an adequate noise bandwidth in outer k-space. A noise tensor is composed over all coil elements and multiplied by its conjugate transpose. Finally, Cholesky decomposition is performed, effectively minimizing noise correlation. 

\begin{figure}[!t]
    \begin{subfigure}{0.33\textwidth}
        \centering
        \includegraphics[width=\textwidth]{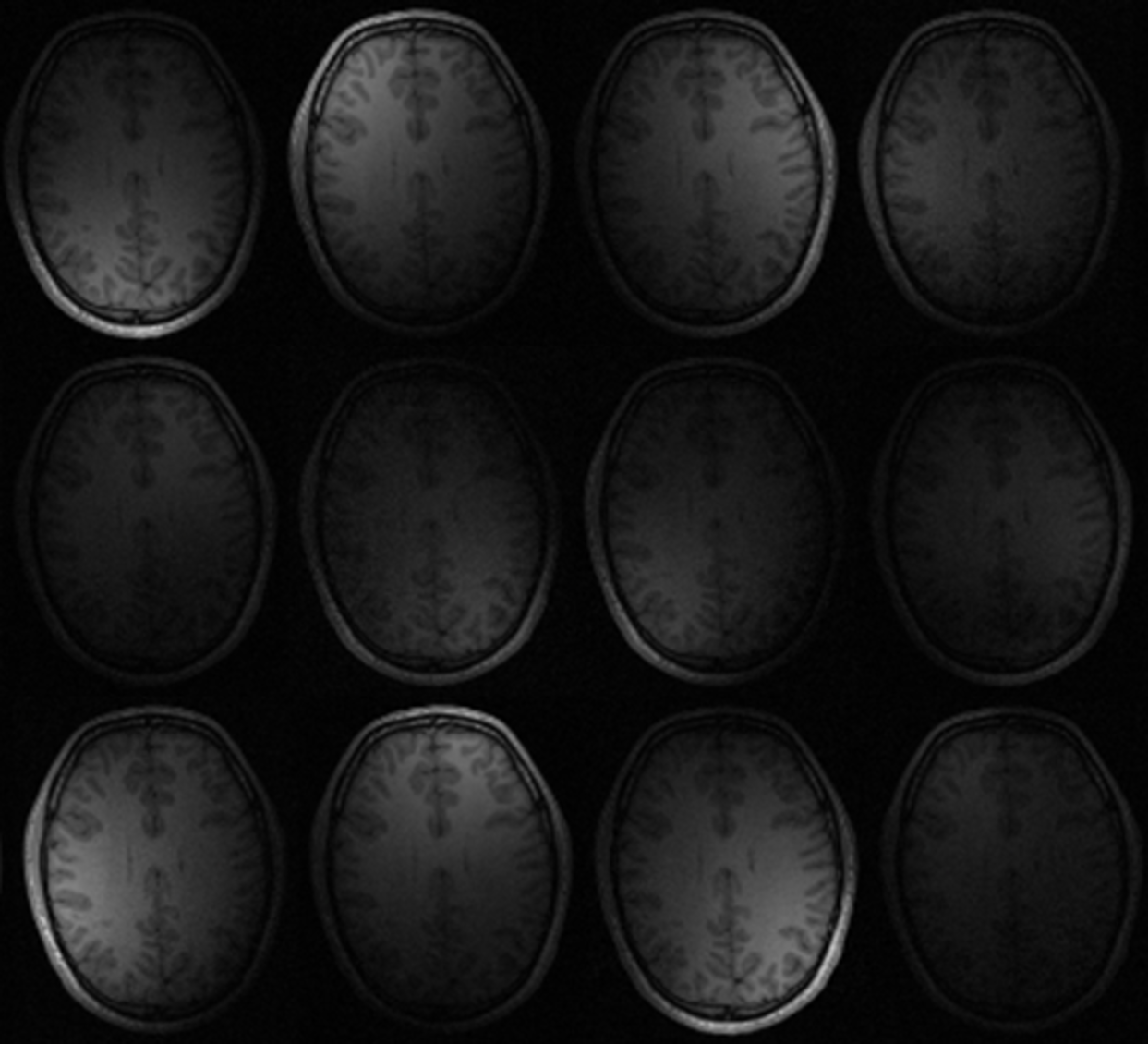}
        \caption{Ground Truth (GT)}
        \label{fig:cc359-12-coils}
    \end{subfigure}
    \hfill
    \begin{subfigure}{0.33\textwidth}
        \centering
        \includegraphics[width=\textwidth]{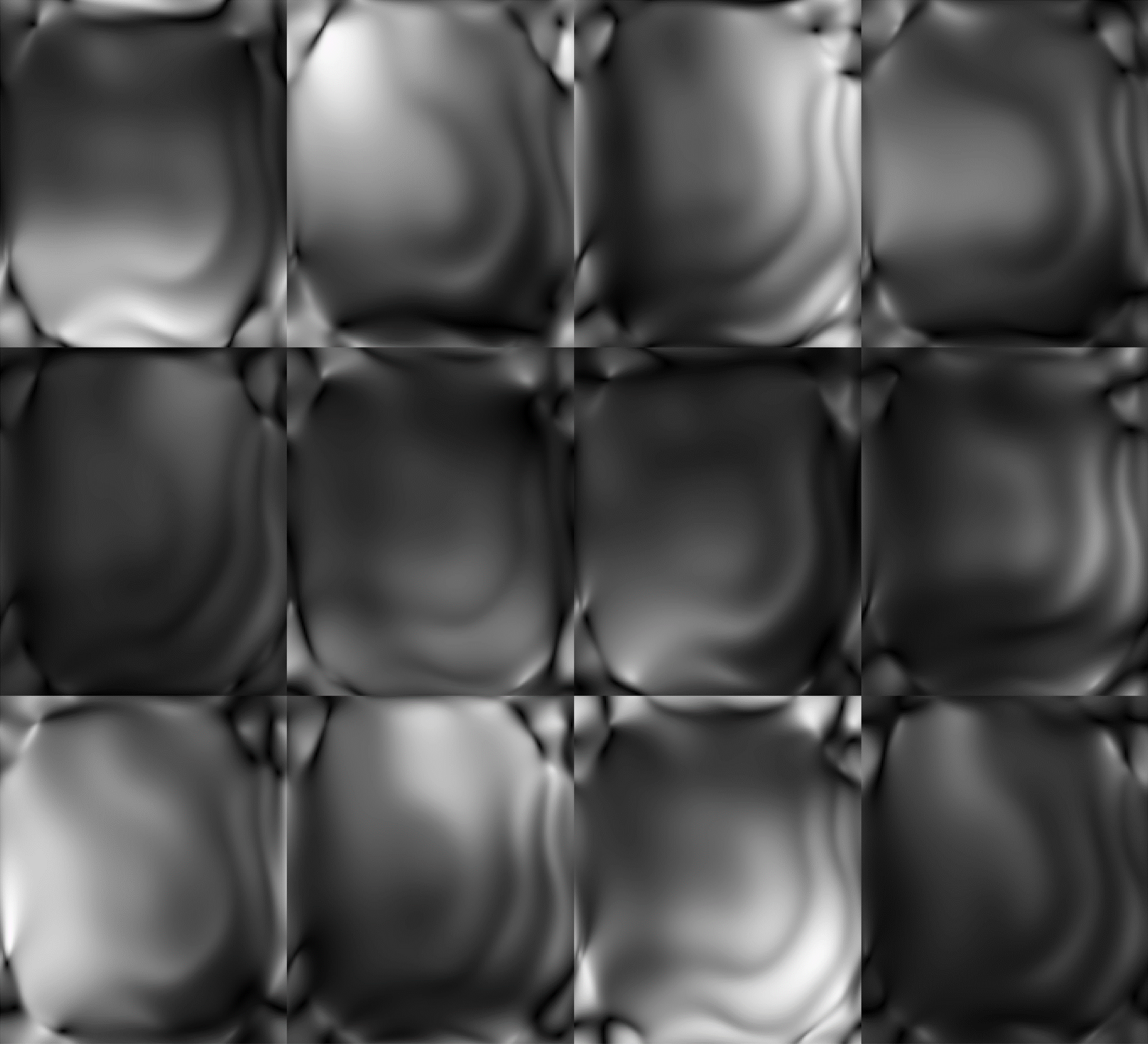}
        \caption{Coil Sensitivity Maps (CSM)}
        \label{fig:cc359-csm-all-coils}
    \end{subfigure}
    \hfill
    \begin{subfigure}{0.33\textwidth}
        \begin{subfigure}{\textwidth}
            \centering
            \includegraphics[width=\textwidth]{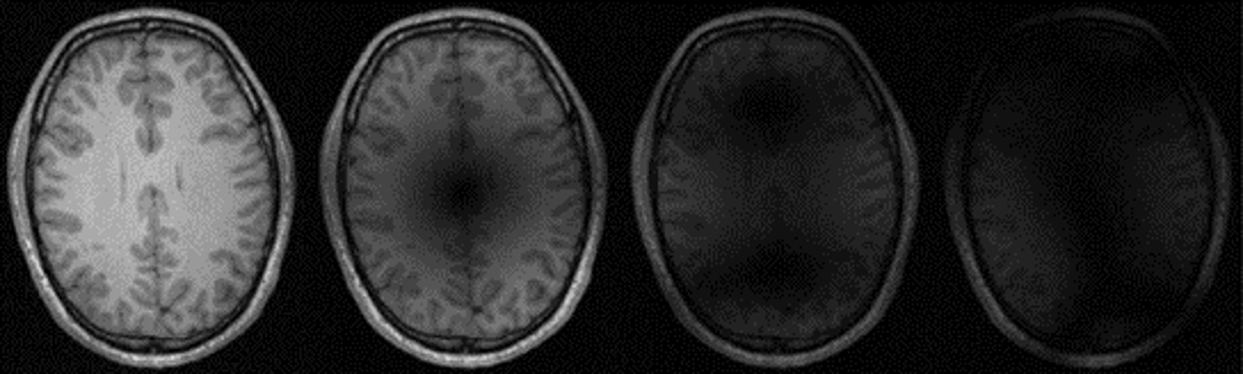}
            \caption{Geometric Decomposition Coil Compression (GDCC)}
            \label{fig:cc359-gdcc}
        \end{subfigure}
        \vspace{0.075\linewidth}
        \vfill
        \begin{subfigure}{\textwidth}
            \centering
            \includegraphics[width=\textwidth]{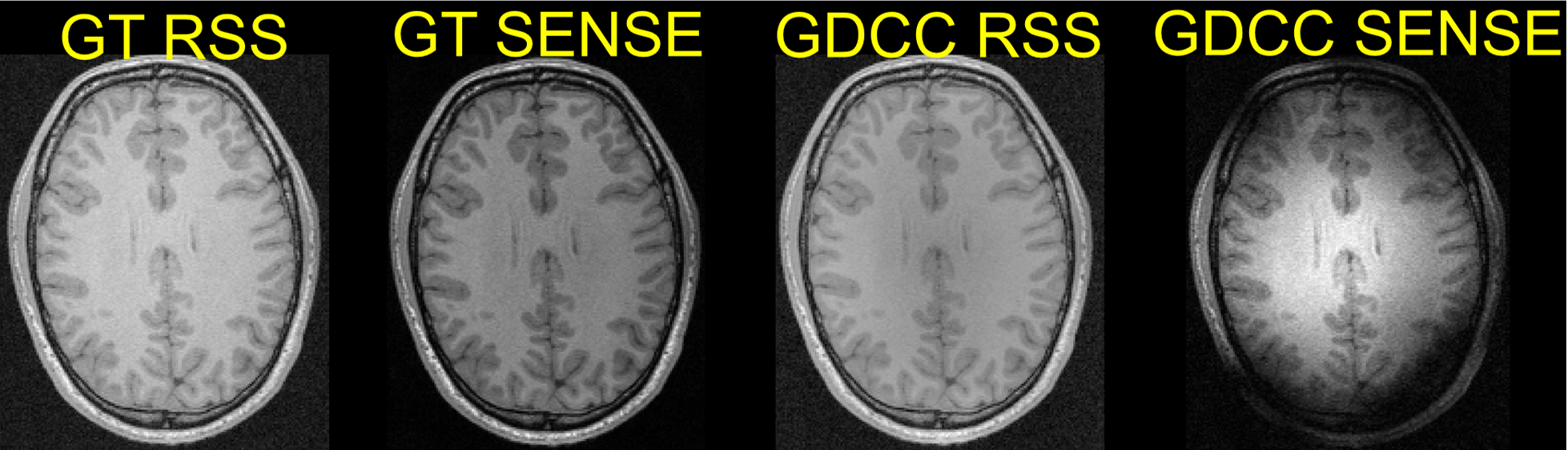}
            \caption{Coil Combination Methods (CCM}
            \label{fig:cc359-gt-rss-sense-gdcc}
        \end{subfigure}
    \end{subfigure}
    \caption{Multicoil-related transforms applied to example data from the CC359 dataset (\cite{beauferrisMultiCoilMRIReconstruction2022}). Fig. \ref{fig:cc359-12-coils} shows the fully sampled 12-coil Ground Truth (GT) data. Fig. \ref{fig:cc359-csm-all-coils} shows the estimated 12-coil Coil Sensitivity Maps (CSM) using the \texttt{EstimateCoilSensitivityMaps} class. In Fig. \ref{fig:cc359-gdcc}, the 12-coil data are reduced to 4 coils after using the Geometric Decomposition Coil Compression (GDCC) method. Fig. \ref{fig:cc359-gt-rss-sense-gdcc} shows different Coil Combination Methods (CSM), such as the Root-Sum-of-Squares (RSS) and the SENsitivity Encoding (SENSE), applied to the Ground Truth (GT) (first and second, respectively), and the GDCC (third and fourth, respectively).}
    \label{fig:coil-transforms}
\end{figure}

The \texttt{GeometricDecompositionCoilCompression} class can compress multicoil data using the geometric decomposition method (\cite{zhangCoilCompressionAccelerated2013}). The \texttt{gcc\_virtual\_coils} parameter defines the number of virtual coils to compress the multicoil data to. The \texttt{gcc\_calib\_lines} parameter is the number of calibration lines used for coil compression. The \texttt{gcc\_align\_data} parameter aligns the data before coil compression. An example of compressing 12-coil data to 4-virtual-coil data, with the \texttt{GeometricDecompositionCoilCompression} transformation is presented in Fig. \ref{fig:cc359-gdcc}.

Cropping in both the image space and k-space may be performed using the \texttt{Cropper} class. Note that when cropping is applied in k-space, the Field-of-View (FOV) changes as a result. When applied in image space, the FOV remains the same while the spatial resolution changes. The \texttt{kspace\_crop} parameter defines whether the cropping is performed in k-space or image space. The \texttt{crop\_before\_masking} parameter defines whether cropping will be applied before or after undersampling the k-space (Sec. \ref{sec:accelerated_mri}). Note that cropping after undersampling alters the relative acceleration factor.

The \texttt{coil\_combination\_method} function allows to perform either Root-Sum-of-Squares (RSS) or SENSE (\cite{pruessmannSENSESensitivityEncoding1999}) coil-combination (Fig. \ref{fig:cc359-gt-rss-sense-gdcc}). In the case of SENSE, coil sensitivity maps need to be available. The \texttt{EstimateCoilSensitivityMaps} class allows estimating coil sensitivity maps on the fly without needing to pre-compute and store them beforehand (Fig. \ref{fig:cc359-csm-all-coils}). The available options are adjusted according to the DIRECT toolkit (\cite{yiasemisDIRECTDeepImage2022}) and include \texttt{ESPIRIT} (\cite{ueckerESPIRiTEigenvalueApproach2014}), Root-Sum-of-Squares (RSS), and unitary methods. Also, the option of estimating coil sensitivity maps using a neural network (UNet) is available by toggling the \texttt{estimate\_coil\_sensitivity\_map\_with\_nn} parameter. This option allows training a model end-to-end (\cite{junJointDeepModelbased2021, sriramEndtoEndVariationalNetworks2020}) and can be combined with the \texttt{EstimateCoilSensitivityMaps} class for optimized coil sensitivity maps estimation.

Normalization can be applied through the \texttt{Normalizer} class. The \texttt{normalize\_inputs} parameter determines whether the inputs will be normalized. The \texttt{normalization\_type} parameter determines the normalization method. The \texttt{minmax} and \texttt{max} methods normalize the data as $\frac{\mathrm{data} - \mathrm{\min(data)}}{\mathrm{\max(data)} - \mathrm{\min(data)}}$ and $\frac{\mathrm{data}}{\mathrm{\max(data)}}$, respectively, in the range 0-1. The \texttt{mean\_std} and the \texttt{mean\_var} methods normalize the data as $\frac{\mathrm{data} - \mathrm{mean(data)}}{\mathrm{std(data)}}$ and $\frac{\mathrm{data} - \mathrm{mean(data)}}{\mathrm{var(data)}}$ respectively. The \texttt{grayscale} method first normalizes the data as in \texttt{minmax} and then multiplies by $255$ to bring the data in the range 0-255. The options \texttt{fft} and \texttt{none} do not apply normalization. When handling complex-valued data, \texttt{fft} can be more intuitive. Finally, the \texttt{kspace\_normalization} parameter determines whether the normalization is performed in k-space, when complex-valued data are available. By default normalization is performed in image space.

Finally, the \texttt{Composer} class allows composing a series of transforms into a single transform. 

\subsection{Training \& Testing Deep Learning models for MRI tasks in ATOMMIC}
\label{sec:dl-models-atommic}

Training and testing DL models in ATOMMIC requires a configuration (YAML) file and a single command to set its path, i.e., \texttt{atommic run -c path\_to\_configuration\_file}. The configuration file allows setting various MRI transforms (as discussed in Sec. \ref{sec:mri-transforms}), undersampling options (as explained in Sec. \ref{sec:accelerated_mri}), and hyperparameters (Fig. \ref{fig:atommic_schematic_overview}). The installation is simple through \texttt{pip install atommic}. Multi-GPU and multi-node training, mixed-precision (floating-point 16), early stopping, and Exponential Moving Average can also be configured. For exporting and logging models, tensorboard\footnote{\url{https://github.com/tensorflow/tensorboard}} and Weights \& Biases\footnote{\url{https://github.com/wandb/wandb}} support is available.

\subsection{Experiments}
\label{sec:experiments}

In our comprehensive evaluation, we demonstrate distinct applications of ATOMMIC in the tasks of accelerated MRI reconstruction, quantitative parameter map estimation, segmentation, and MTL for joint reconstruction and segmentation. Twenty-five DL models were benchmarked in various public datasets and with different hyperparameters, as presented in Table \ref{tab:comparative_analysis_atommic}.

\subsubsection{Datasets}
\label{sec:datasets}

For the task of accelerated MRI reconstruction, three datasets were used: the Calgary Campinas 359 dataset (CC359) (\cite{beauferrisMultiCoilMRIReconstruction2022}), the fastMRI Brains multicoil dataset (fastMRIBrains) (\cite{zbontarFastMRIOpenDataset2019}), and the Stanford Fully Sampled 3D FSE Knee dataset (StanfordKnee) (\cite{eppersonCreationFullySampled}). 

The CC359 dataset comprises 117 3D T$_1$ weighted twelve-coil brain scans. The size of the acquisition matrix is 256 $\times$ 218, while it varies in the slice-encoding (kz) direction between 170 and 180 due to 15\% zero-filling. In our experiments, every subject's first 50 and last 50 slices were excluded since these mainly resided outside the brain. The training set consisted of 47 subjects, the validation set of 20 subjects, and the test set of 50 subjects. A 2D Poisson disc distribution sampling pattern accelerated imaging by 5x and 10x times. 

The fastMRIBrains dataset comprises of T$_1$-weighted, T$_1$-weighted with contrast agent (T$_1$POST), T$_2$-weighted, and Fluid-Attenuated Inversion Recovery (FLAIR) scans. In our experiments, we used the first batch of data out of 10, which contained 449 subjects in the training set and 457 subjects in the validation set. Nine subjects were removed due to containing not-a-number (NaN) values. The number of coils varied from four to twenty. The matrix size ranged from minimum 512 to maximum 768 $\times$ minimum 213 to maximum 396 and was cropped to 320 $\times$ minimum 213 to maximum 320. An Equispaced 1D sampling pattern accelerated imaging by 4x and 8x times. 

The StanfordKnee dataset consists of Proton-Density (PD) 3D Fast-Spin Echo (FSE) eight-coil data with fat saturation. The dataset included 19 subjects, split into 13 subjects for the training set, 3 for the validation set, and 3 for the test set. The matrix size was 320 $\times$ 320 $\times$ 256. A Gaussian 2D sampling pattern was used to accelerate imaging by 12x times.

The Amsterdam Ultra-high field adult lifespan database (AHEAD) (\cite{alkemadeAmsterdamUltrahighField2020}) dataset was used to estimate quantitative parameter maps, as it contains multi-echo data necessary for this task. It consists of thirty-two-coil $T_{1}$, $T_{2^{*}}$ and Quantitative Susceptibility Mapping brain scans of four echo times MP2RAGE-ME 7 Tesla (\cite{caanMP2RAGEMET1T22019}). Motion correction with Fat navigators (FatNavs) and defacing in the image domain was already applied to the dataset (\cite{zhangUnifiedModelReconstruction2022}). The scanned image resolution is 0.7mm isotropic. The objective was to estimate the following quantitative maps: $R^*_{2}$, $B_0$, and the angle of the net magnetization $M$, denoted as $|M|$. We used the first ten subjects of the dataset, 001 to 010, of which the first six were used for training, the next two for validation, and the last two for testing. A Gaussian 2D sampling pattern was used to accelerate imaging by 12x times. Brain tissue segmentation masks were pre-computed and applied during training to avoid including NaN or infinity (Inf) values on the skull or the background. Brain tissue masks were computed by applying Otsu's thresholding, computing the largest connected component and the convex hull, and applying a series of binary erosions and dilations.

For the segmentation task, three datasets were used: the Brain Tumor Segmentation 2023 Adult Glioma challenge dataset (BraTS2023AdultGlioma) (\cite{kazerooniBrainTumorSegmentation2024}), the Ischemic Stroke Lesion Segmentation 2022 (ISLES2022SubAcuteStroke) challenge dataset (\cite{hernandezpetzscheISLES2022Multicenter2022}), and the segmentation-only dataset of the Stanford Knee MRI with Multi-Task Evaluation (SKM-TEA) dataset (\cite{desaiSKMTEADatasetAccelerated2022}). 

The BraTS2023AdultGlioma dataset contains 1000 subjects on the training set and 251 on the validation set, while no ground truth test labels are available. The objective is to segment four classes: necrotic tumor core, peritumoral edematous/invaded tissue, gadolinium-enhancing tumor, and whole tumor. The ISLES2022SubAcuteStroke dataset includes Apparent Diffusion Coefficient (ADC) maps, FLAIR scans, and Diffusion Weighted Imaging (DWI) scans. The training set consisted of 172 subjects, the validation set 37, and the test set 38. The objective is to segment one class, specified as stroke lesions. 

The SKM-TEA dataset contains complex-valued multicoil raw data, real-valued coil-combined data, and ground truth segmentation labels, allowing for both segmentation independently and MTL for combined reconstruction and segmentation. Data are of heterogeneous patient anatomy with potential distribution shifts being present as data were acquired from multiple vendors (\cite{desaiSKMTEADatasetAccelerated2022}). The SKM-TEA segmentation-only dataset provides data imaged in the sagittal plane, with four segmentation classes: lateral tibial cartilage, medial tibial cartilage, lateral meniscus, and medial meniscus. In contrast, for MTL, the complex-valued SKM-TEA dataset comprises data reconstructed in the axial plane with both phase-encoding dimensions. The data are stored as $x \times ky \times kz$, where $x$ denotes the number of slices, and $ky \times kz$, the dimensions to apply the provided undersampling mask and coil sensitivity maps. The matrix size is 512 $\times$ 160 and is cropped to 416 $\times$ 80 to remove oversampling. Data are undersampled using a Poisson disc distribution 2D pattern with an acceleration factor 4x. Both for segmentation only and for MTL, we split the SKM-TEA dataset into 86 subjects in the training set, 33 in the validation set, and 36 in the test set.

\subsubsection{Hyperparameters}
\label{sec:hyperparameters}

For the tasks of reconstruction and quantitative parameter map estimation, models were trained for 20 epochs. For estimating quantitative parameter maps from an accelerated MRI acquisition, we first trained reconstruction models and then used them to initialize the quantitative parameter map estimation models. For the task of segmentation, models were trained for 20 epochs on the BraTS2023AdultGlioma and the SKM-TEA segmentation-only datasets and for 50 epochs on the ISLES2022SubAcuteStroke dataset since its size was significantly smaller than the other two. For MTL for joint reconstruction and segmentation, models were trained for 15 epochs. 

The learning rate was set to $10^{-4}$, and the floating point precision was set to mixed 16 for all models on all tasks. Normalization by the max value (Sec. \ref{sec:mri-transforms}) was applied to all models trained for reconstruction, segmentation, and MTL. To stabilize the training of quantitative parameter map estimation models on the AHEAD dataset, we heuristically scaled the input multi-echo k-space data by a factor of $10^4$ and the input quantitative maps by a factor of $10^{-3}$ as in (\cite{zhangUnifiedModelReconstruction2022}). The AHEAD data consist of four echo times, for which the values were 3 ms, 11.5 ms, 20 ms, and 28.5 ms, respectively. 

Models were trained and tested on an Nvidia Tesla V100 GPU with 32GB memory. A detailed overview of the selected hyperparameters for each model is presented in the Appendix (Table \ref{tab:hp-overview}). Trained models' checkpoints are available on HuggingFace\footnote{\url{https://huggingface.co/wdika}}. 

\begin{table*}[!t]
\centering
\caption{Comparative evaluation of DL models using ATOMMIC for different MRI tasks. The first column reports the task, specifically MultiTask Learning (MTL) (second row) for jointly performing accelerated MRI reconstruction (REC) (fourth row) and MRI segmentation (SEG) (fifth row), and quantitative MRI (qMRI) for estimating parameter maps (second row). The second column reports the publicly available datasets used for training and testing. The third column reports the coil sensitivity maps (CSM) estimation method and the coil combination method (CCM) (Sec. \ref{sec:mri-transforms}). When CSMs were not available, they were estimated with the \texttt{EstimateCoilSensitivityMaps} transformation of ATOMMIC or End-to-End during training with a UNet. The \texttt{GeometricDecompositionCoilCompression} (GDCC) transformation was applied to the fastMRI Brains Multicoil dataset, reducing various coils (four to twenty) into single coil data. The CCM was set either to Sensitivity Encoding (SENSE) or Root-Sum-of-Squares (RSS). The fourth column reports the used optimizer (Opt) and learning rate scheduler (LRS). For Optim, the Adam and Adam weighted (AdamW) were used, and for LR Sched, the Inverse Square Root Annealing (ISRA) and Cosine Annealing (CA) were used. The used loss function is reported in the fifth column, and the trained and tested DL models are reported in the sixth column.}
\mediumsize % Reducing the font size
\begin{tabularx}{\textwidth}{p{0.5cm}p{2.2cm}p{0.5cm}p{0.5cm}p{2.55cm}p{8cm}}
    \multicolumn{1}{c}{\textbf{Task}} & \multicolumn{1}{c}{\textbf{Dataset}} & \multicolumn{1}{c}{\textbf{CSM-CCM}} & \multicolumn{1}{c}{\textbf{Opt-LRS}} & \multicolumn{1}{c}{\textbf{Loss}} & \multicolumn{1}{c}{\textbf{Models}}  \\
    \hline
    MTL & 
    \begin{tabular}{p{2.3cm}}
        SKM-TEA (\cite{desaiSKMTEADatasetAccelerated2022})
    \end{tabular} & 
    \begin{tabular}{p{1.5cm}}
        Available -SENSE
    \end{tabular} &
    \begin{tabular}{p{1.5cm}}
        Adam-ISRA
    \end{tabular} &
    \begin{tabular}{p{2.5cm}}
        0.5*L1 + 0.5*DICE
    \end{tabular} &
    \begin{tabular}{p{5.5cm}}
        Image domain Deep Structured Low-Rank Network (IDSLR) (\cite{pramanikRECONSTRUCTIONSEGMENTATIONPARALLEL2021}) \\
        Image domain Deep Structured Low-Rank UNet (IDSLRUNet) (\cite{pramanikRECONSTRUCTIONSEGMENTATIONPARALLEL2021}) \\
        Multi-Task Learning for MRI Reconstruction and Segmentation (MTLRS) \cite{karkalousosMultiTaskLearningAcceleratedMRI2024}) \\
        Segmentation Network MRI (SegNet) (\cite{sunJointCSMRIReconstruction2019})
    \end{tabular} \\
    \hline
    qMRI & 
    \begin{tabular}{p{2.3cm}}
        AHEAD (\cite{alkemadeAmsterdamUltrahighField2020})
    \end{tabular} & 
    \begin{tabular}{p{1.5cm}}
        Available -SENSE
    \end{tabular} &
    \begin{tabular}{p{1.5cm}}
        Adam-ISRA
    \end{tabular} &
    \begin{tabular}{p{2.5cm}}
        SSIM
    \end{tabular} &
    \begin{tabular}{p{5.5cm}}
        quantitative Cascades of Independently Recurrent Inference Machines (qCIRIM) \\
        quantitative End-to-End Variational Network (qVarNet) (\cite{zhangUnifiedModelReconstruction2022})
    \end{tabular} \\
    \hline
    REC & 
    \begin{tabular}{p{2.3cm}}
        CC359 (\cite{beauferrisMultiCoilMRIReconstruction2022}) \\ \\ \\ \\ \\ \\
        fastMRI Brains Multicoil (\cite{zbontarFastMRIOpenDataset2019}) \\ \\ \\
        Stanford Knees (\cite{eppersonCreationFullySampled})
    \end{tabular} & 
    \begin{tabular}{p{1.5cm}}
        End-to-End -RSS \\ \\
        GDCC -SENSE \\ \\ \\
        ATOMMIC -SENSE
    \end{tabular} & 
    \begin{tabular}{p{1.5cm}}
        AdamW-CA \\ \\ \\ \\
        Adam-ISRA \\ \\ \\ \\
        AdamW-ISRA \\
    \end{tabular} &
    \begin{tabular}{p{2.5cm}}
        0.9*SSIM + 0.1*L1 \\ \\ \\ \\ \\
        0.9*SSIM + 0.1*L1 \\ \\ \\ \\ \\
        Wasserstein (\cite{cuturiSinkhornDistancesLightspeed2013}) \\
    \end{tabular} &
    \begin{tabular}{p{5.5cm}}
        Cascades of Independently Recurrent Inference Machines (CIRIM) (\cite{karkalousosAssessmentDataConsistency2022}) \\
        Convolutional Recurrent Neural Networks (CRNN) (\cite{qinConvolutionalRecurrentNeural2019}) \\
        Deep Cascade of Convolutional Neural Networks (CascadeNet) (\cite{schlemperDeepCascadeConvolutional2018}) \\
        End-to-End Variational Network (VarNet) (\cite{sriramEndtoEndVariationalNetworks2020}) \\
        Joint Deep Model-Based MR Image and Coil Sensitivity Reconstruction Network (JointICNet) (\cite{junJointDeepModelbased2021}) \\
        KIKINet (\cite{eoKIKInetCrossdomainConvolutional2018}) \\
        Learned Primal-Dual Net (LPDNet) (\cite{adlerLearnedPrimalDualReconstruction2018}) \\
        Model-based Deep Learning Reconstruction (MoDL) (\cite{aggarwalMoDLModelBased2019}) \\
        Recurrent Inference Machines (RIM) (\cite{lonningRecurrentInferenceMachines2019}) \\
        Recurrent Variational Network (RVN) (\cite{yiasemisRecurrentVariationalNetwork2022}) \\
        UNet (\cite{ronnebergerUNetConvolutionalNetworks2015}) \\
        Variable Splitting Network (VSNet) (\cite{duanVSNetVariableSplitting2019}) \\
        XPDNet (\cite{ramziBenchmarkingMRIReconstruction2020})
    \end{tabular} \\
    \hline
    SEG & 
    \begin{tabular}{p{2.3cm}}
        BraTS 2023 Adult Glioma (\cite{kazerooniBrainTumorSegmentation2024}) \\ \\
        ISLES 2022 Sub Acute Stroke (\cite{hernandezpetzscheISLES2022Multicenter2022}) \\ \\ \\
        SKM-TEA (\cite{desaiSKMTEADatasetAccelerated2022}) \\
    \end{tabular} & 
    &
    \begin{tabular}{p{1.5cm}}
        AdamW-ISRA \\ \\ \\
        Adam-CA \\ \\ \\ \\
        AdamW-ISRA \\
    \end{tabular} &
    \begin{tabular}{p{2.5cm}}
        DICE \\ \\ \\ \\
        DICE \\ \\ \\ \\ \\
        DICE \\
    \end{tabular} &
    \begin{tabular}{p{5.5cm}}
        Attention UNet (\cite{oktayAttentionUNetLearning2018}) \\
        Dynamic UNet (DYNUNet) (\cite{isenseeNnUNetSelfconfiguringMethod2021}) \\
        UNet 2D (\cite{ronnebergerUNetConvolutionalNetworks2015}) \\
        UNet 3D (\cite{ronnebergerUNetConvolutionalNetworks2015}) \\
        VNet (\cite{milletariVNetFullyConvolutional2016})
    \end{tabular}
\end{tabularx}
\label{tab:comparative_analysis_atommic}
\end{table*}

\subsubsection{Evaluation metrics}
\label{sec:eval-metrics}

The performance in the reconstruction task (CC359, fastMRIBrains, and StanfordKnee datasets), in the task of reconstruction and quantitative parameter map estimation (AHEAD dataset), and in the task of reconstruction for MTL (SKM-TEA dataset) was evaluated by measuring the similarity of the predicted reconstructions and the ground truth images using the Structural Similarity Index Measurement (SSIM) (\cite{wangImageQualityAssessment2004}) and assessing the perceived image quality using the Peak Signal-to-Noise Ratio (PSNR). For evaluating the performance of quantitative parameter map estimation models, the Normalized Mean Squared Error (NMSE) was also computed.

The accuracy of the segmentation models in the BraTS2023AdultGlioma and the SKM-TEA datasets was evaluated by quantifying the similarity between the predicted segmentations and the ground truth labels, measured by the DICE coefficient and the Intersection over Union (IOU). The accuracy of segmentation boundaries was assessed by computing the Hausdorff Distance 95\% (HD95), which provides insights into the largest segmentation errors while minimizing the influence of outliers. Additionally, the significance of false positives and false negatives was measured by the F1 score. 

Different metrics were used to evaluate the performance of segmentation models on the ISLES2022SubAcuteStroke dataset, as specified in the challenge manuscript (\cite{hernandezpetzscheISLES2022Multicenter2022}). Specifically, due to small lesions, such as punctiform infarcts, an increase in the DICE coefficient might result from detecting only a large lesion. Therefore, the Absolute Volume Difference (AVD) was used to compute voxel-wise differences in the infarct volume, while lesion-wise metrics such as the Absolute Lesion Difference (ALD) and the Lesion F1 (L-F1) score allowed measuring of the lesion detection and to count the lesion burden accurately. 

\section{Results}
\label{sec:results}

Table \ref{tab:recon_cc359_fastMRIBrains} presents the reconstruction task performance of models trained on the CC359 and fastMRIBrains datasets. The Variational Network (VarNet) achieved the highest SSIM and PSNR scores for 5x acceleration and the highest PSNR score for 10x acceleration on the CC359 dataset. The Joint Deep Model-Based MR Image and Coil Sensitivity Reconstruction Network (JointICNet) scored the highest SSIM for 10x acceleration on the CC359 dataset. In contrast, on the fastMRIBrains dataset, the Recurrent Variational Network (RVN) scored the highest SSIM and PSNR scores for 4x acceleration and the VarNet for 8x acceleration. The Cascades of Independently Recurrent Inference Machines (CIRIM) yielded the highest SSIM and PSNR scores on the StanfordKnee dataset for 12x acceleration, as presented in Table \ref{tab:recon_stanford_knees}. Conversely, on the same dataset, the Convolutional Recurrent Neural Network (CRNN) was excluded from the analysis due to unstable gradient computation, although trained across a wide range of learning rates ($10^{-4}$ to $10^{-9}$).

\begin{table*}[!ht]
    \centering
    \caption{Overview of performance on reconstructing accelerated brain data. In the first column, the name of the model is reported. The rest of the columns report SSIM and PSNR scores for each dataset, where up arrows indicate the highest the best. The second to fifth columns report the performance of each model on the CC359 dataset for 5x (second-third columns) \& 10x (fourth and fifth column) Poisson 2D undersampling. The sixth to ninth columns report performance on the fastMRIBrain dataset, for 4x (sixth and seventh column) \& 8x (eighth and ninth columns) Equispaced 1D acceleration. Best performing models are highlighted in bold. Methods are sorted in alphabetical order.}
    \begin{adjustbox}{width=\textwidth}
    \begin{tabular}{l|cccc|cccc}
        & \multicolumn{4}{c|}{CC359 - Poisson 2D} & \multicolumn{4}{c}{fastMRIBrains - Equispaced 1D} \\
        Model & \multicolumn{2}{c}{5x} & \multicolumn{2}{c|}{10x} & \multicolumn{2}{c}{4x} & \multicolumn{2}{c}{8x} \\
        & SSIM $\uparrow$ & PSNR $\uparrow$ & SSIM $\uparrow$ & PSNR $\uparrow$ & SSIM $\uparrow$ & PSNR $\uparrow$ & SSIM $\uparrow$ & PSNR $\uparrow$ \\
        \midrule
        CCNN            & 0.845     $\pm$ 0.064      & 28.36     $\pm$ 3.69      & 0.783     $\pm$ 0.089      & 25.95     $\pm$ 3.64      & 0.886     $\pm$ 0.192      & 33.47     $\pm$ 5.92      & 0.836     $\pm$ 0.202      & 29.40     $\pm$ 5.71 \\ 
        CIRIM           & 0.858     $\pm$ 0.074      & 28.79     $\pm$ 4.23      & 0.816     $\pm$ 0.094      & 26.92     $\pm$ 4.36      & 0.892     $\pm$ 0.184      & 33.83     $\pm$ 6.11      & 0.846     $\pm$ 0.202      & 30.23     $\pm$ 5.67 \\ 
        CRNN            & 0.774     $\pm$ 0.088      & 25.59     $\pm$ 4.19      & 0.722     $\pm$ 0.088      & 24.48     $\pm$ 3.39      & 0.868     $\pm$ 0.195      & 31.31     $\pm$ 5.46      & 0.806     $\pm$ 0.198      & 27.50     $\pm$ 5.57 \\ 
        \bf{JointICNet} & 0.872     $\pm$ 0.065      & 29.28     $\pm$ 3.99      & \bf{0.828 $\pm$ 0.086}     & 27.36     $\pm$ 4.10      & 0.832     $\pm$ 0.198      & 28.57     $\pm$ 5.50      & 0.772     $\pm$ 0.202      & 25.50     $\pm$ 5.38 \\ 
        KIKINet         & 0.788     $\pm$ 0.087      & 25.43     $\pm$ 4.16      & 0.742     $\pm$ 0.105      & 24.37     $\pm$ 3.88      & 0.856     $\pm$ 0.201      & 31.02     $\pm$ 5.68      & 0.805     $\pm$ 0.207      & 27.78     $\pm$ 5.82 \\ 
        LPDNet          & 0.849     $\pm$ 0.075      & 28.26     $\pm$ 4.22      & 0.810     $\pm$ 0.099      & 26.73     $\pm$ 4.23      & 0.882     $\pm$ 0.201      & 32.60     $\pm$ 6.78      & 0.840     $\pm$ 0.208      & 29.51     $\pm$ 5.93 \\ 
        MoDL            & 0.844     $\pm$ 0.068      & 27.97     $\pm$ 4.20      & 0.793     $\pm$ 0.088      & 25.89     $\pm$ 4.39      & 0.870     $\pm$ 0.188      & 31.44     $\pm$ 5.66      & 0.813     $\pm$ 0.192      & 27.81     $\pm$ 5.86 \\ 
        RIM             & 0.834     $\pm$ 0.077      & 27.45     $\pm$ 4.32      & 0.788     $\pm$ 0.091      & 25.56     $\pm$ 3.96      & 0.886     $\pm$ 0.188      & 33.12     $\pm$ 6.04      & 0.837     $\pm$ 0.199      & 29.49     $\pm$ 5.74 \\ 
        \bf{RVN}        & 0.845     $\pm$ 0.067      & 28.14     $\pm$ 3.53      & 0.787     $\pm$ 0.093      & 26.03     $\pm$ 3.77      & \bf{0.894 $\pm$ 0.180}     & \bf{34.23 $\pm$ 5.97}     & 0.843     $\pm$ 0.195      & 30.08     $\pm$ 5.68 \\ 
        UNet            & 0.849     $\pm$ 0.070      & 28.85     $\pm$ 4.17      & 0.810     $\pm$ 0.091      & 27.20     $\pm$ 4.20      & 0.885     $\pm$ 0.182      & 33.09     $\pm$ 6.02      & 0.847     $\pm$ 0.197      & 29.87     $\pm$ 5.68 \\ 
        \bf{VarNet}     & \bf{0.874 $\pm$ 0.061}     & \bf{29.49 $\pm$ 3.86}     & 0.827     $\pm$ 0.087      & \bf{27.51 $\pm$ 4.01}     & 0.892     $\pm$ 0.198      & 34.00     $\pm$ 6.30      & \bf{0.856 $\pm$ 0.216}     & \bf{30.73 $\pm$ 5.94} \\ 
        VSNet           & 0.788     $\pm$ 0.079      & 25.51     $\pm$ 3.91      & 0.740     $\pm$ 0.089      & 24.19     $\pm$ 3.27      & 0.856     $\pm$ 0.196      & 30.37     $\pm$ 5.34      & 0.796     $\pm$ 0.197      & 26.88     $\pm$ 5.43 \\ 
        XPDNet          & 0.761     $\pm$ 0.100      & 24.27     $\pm$ 4.14      & 0.700     $\pm$ 0.112      & 22.65     $\pm$ 3.22      & 0.854     $\pm$ 0.212      & 31.03     $\pm$ 6.75      & 0.788     $\pm$ 0.218      & 26.96     $\pm$ 6.18 \\ 
        ZeroFilled      & 0.679     $\pm$ 0.103      & 19.89     $\pm$ 7.45      & 0.656     $\pm$ 0.092      & 19.24     $\pm$ 7.37      & 0.671     $\pm$ 0.194      & 24.12     $\pm$ 6.21      & 0.591     $\pm$ 0.213      & 21.03     $\pm$ 5.97 \\
        \end{tabular}
    \end{adjustbox}
    \label{tab:recon_cc359_fastMRIBrains}
\end{table*}

\begin{table}[!ht]
    \caption{Overview of performance on reconstructing accelerated knee data from the StanfordKnees dataset for 12x Gaussian 2D acceleration. In the first column, the name of the model is reported. The second and third columns report SSIM and PSNR scores, where up arrows indicate the highest the best. Best performing models are highlighted in bold. Methods are sorted in alphabetical order.}
    \label{tab:recon_stanford_knees}
    \centering
    \begin{adjustbox}{width=0.4\textwidth}
        \begin{tabular}{ccc}
            Model & \multicolumn{2}{c}{StanfordKnees - Gaussian 2D 12x} \\
            & SSIM $\uparrow$ & PSNR $\uparrow$ \\
            \midrule
            CCNN       & 0.767 $\pm$ 0.299      & 31.64 $\pm$ 6.84 \\ 
            \textbf{CIRIM} & \textbf{0.795 $\pm$ 0.311} & \textbf{32.76 $\pm$ 7.20} \\ 
            JointICNet & 0.728 $\pm$ 0.291      & 29.59 $\pm$ 6.31 \\ 
            KIKINet    & 0.659 $\pm$ 0.241      & 27.35 $\pm$ 5.54 \\ 
            LPDNet     & 0.736 $\pm$ 0.297      & 29.75 $\pm$ 6.31 \\ 
            MoDL       & 0.566 $\pm$ 0.284      & 23.63 $\pm$ 4.66 \\ 
            RIM        & 0.769 $\pm$ 0.304      & 31.58 $\pm$ 6.74 \\ 
            RVN        & 0.778 $\pm$ 0.301      & 31.96 $\pm$ 7.00 \\ 
            UNet       & 0.771 $\pm$ 0.296      & 31.37 $\pm$ 6.54 \\ 
            VarNet     & 0.764 $\pm$ 0.302      & 31.48 $\pm$ 6.73 \\ 
            VSNet      & 0.708 $\pm$ 0.289      & 28.47 $\pm$ 5.82 \\ 
            XPDNet     & 0.654 $\pm$ 0.270      & 27.16 $\pm$ 5.81 \\ 
            ZeroFilled & 0.548 $\pm$ 0.196      & 18.07 $\pm$ 6.20 \\
        \end{tabular}
    \end{adjustbox}
\end{table}

\begin{figure}[!ht]
    \centering
    \begin{subfigure}[t]{0.485\textwidth}
        \includegraphics[width=\textwidth]{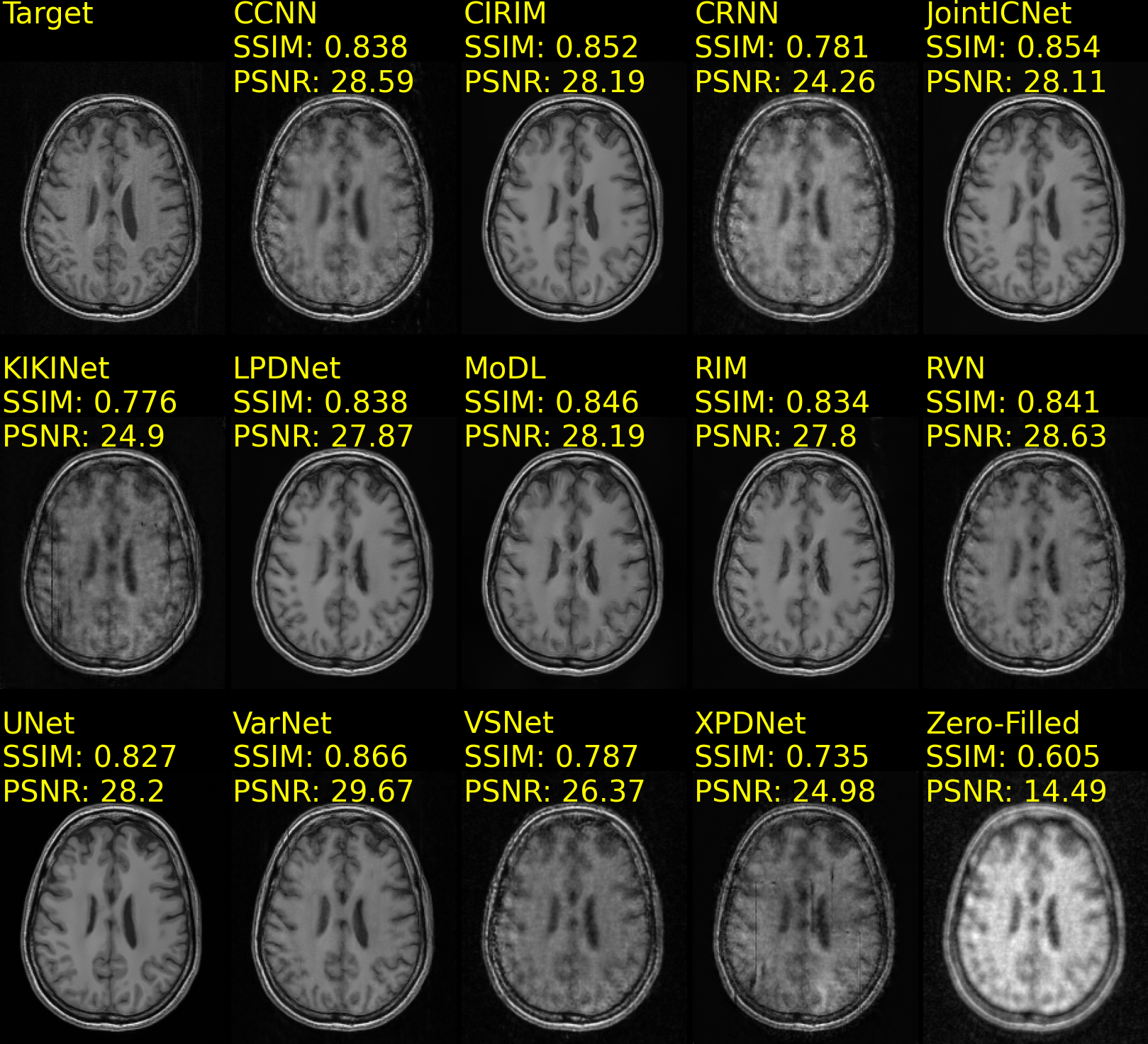}
        \caption{CC359 12-coil data - 5x acceleration}
        \label{subfig:recon_cc359-5x}
    \end{subfigure}
    \begin{subfigure}[t]{0.485\textwidth}
        \includegraphics[width=\textwidth]{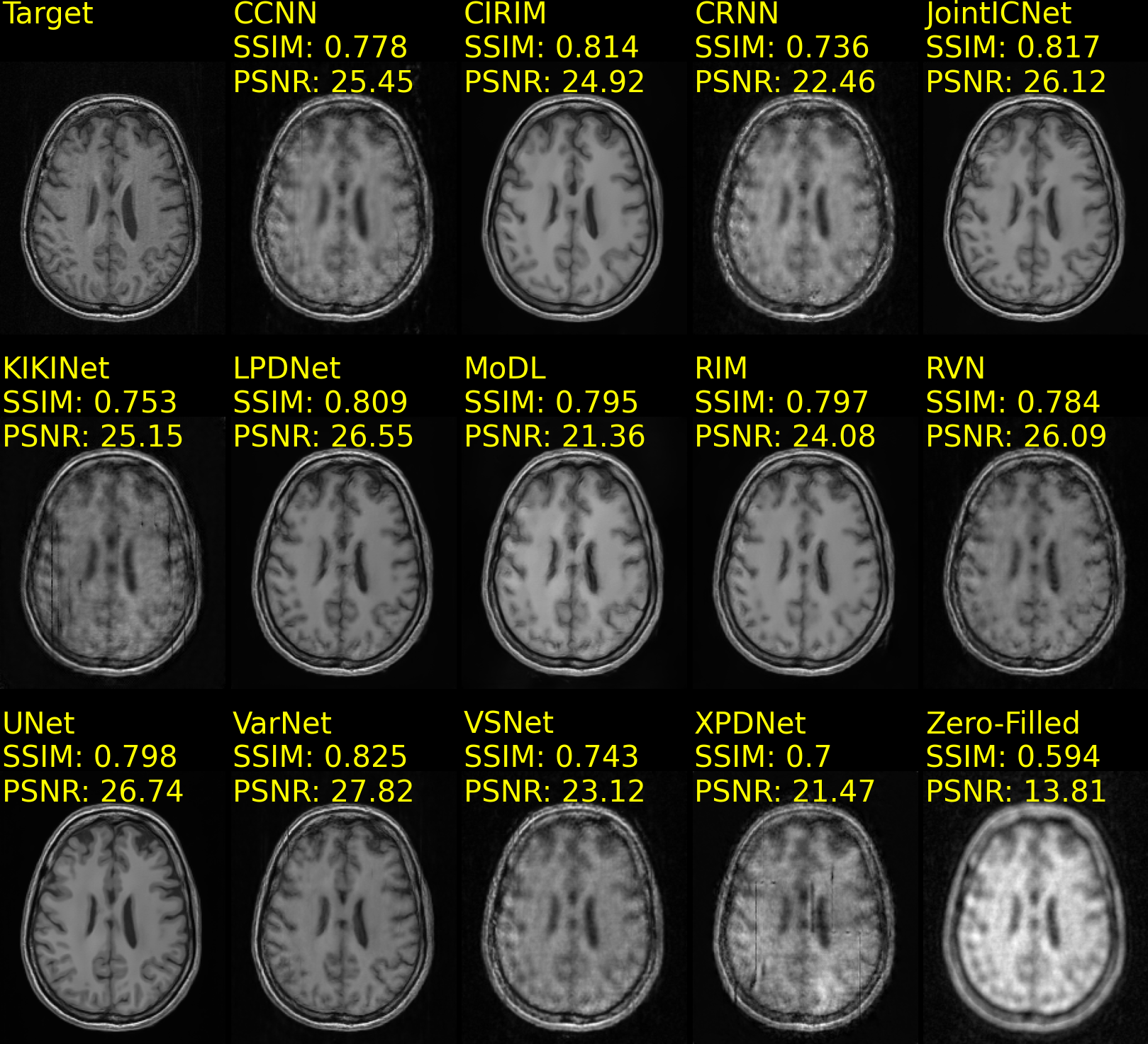}
        \caption{CC359 12-coil data - 10x acceleration}
        \label{subfig:recon_cc359-10x}
    \end{subfigure}
    \caption{Reconstructions of 12-coil T$_1$-weighted data from the CC359 dataset, undersampled with a Poisson disc distribution 2D sampling pattern for 5x (Fig. \ref{subfig:recon_cc359-5x}) and 10x (Fig. \ref{subfig:recon_cc359-10x}) acceleration. The top row-first column shows the ground truth (Target) image. SSIM and PSNR scores are reported for each method and computed against the Target image. Methods are sorted alphabetically.}
    \label{fig:recon_cc359}
\end{figure}

\begin{figure}[!ht]
    \centering
    \begin{subfigure}[t]{0.485\textwidth}
        \includegraphics[width=\textwidth]{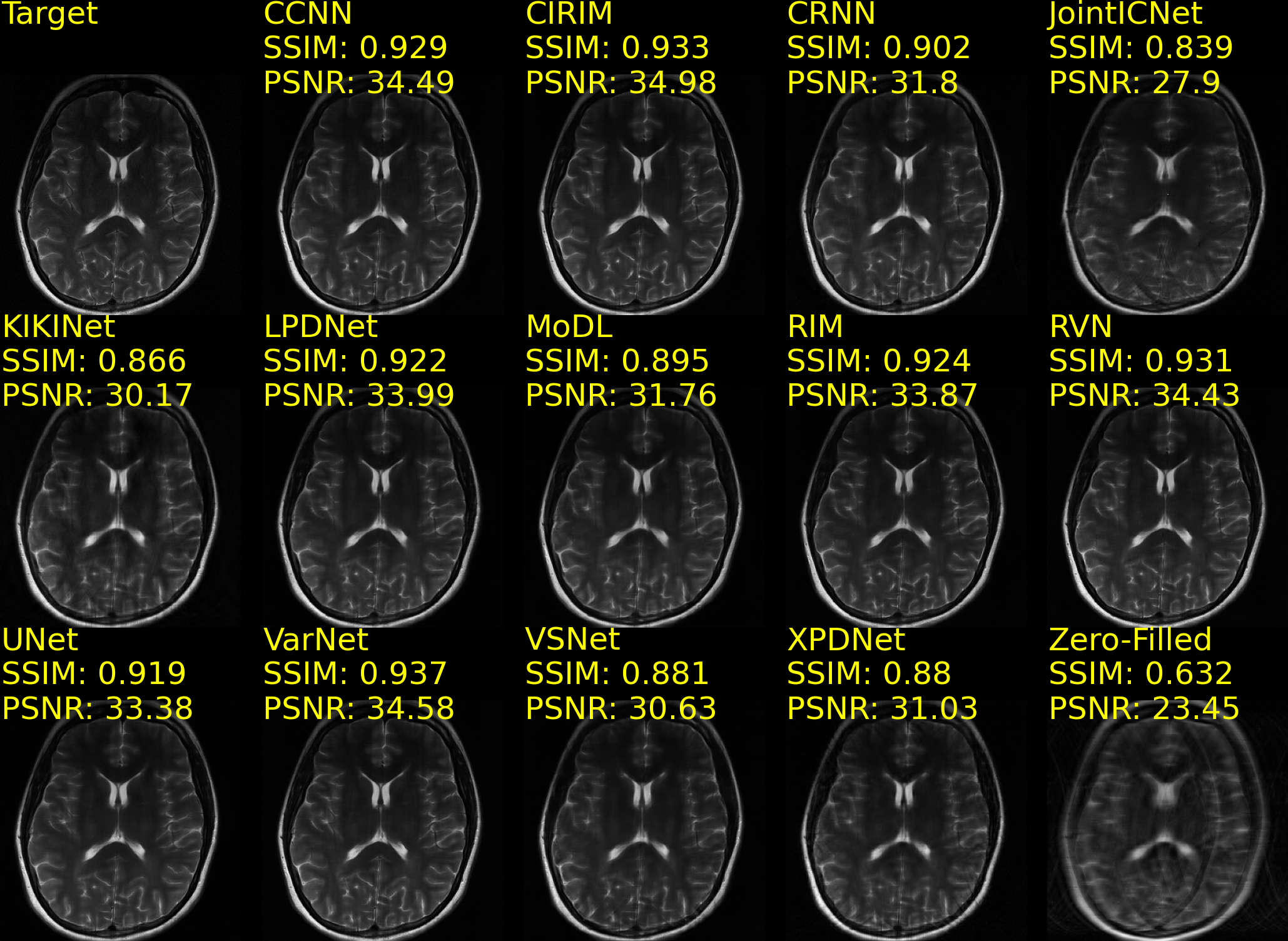}
        \caption{fastMRIBrains 14-coil data - 4x acceleration}
        \label{subfig:recon_fastmri_brains-4x}
    \end{subfigure}
    \begin{subfigure}[t]{0.485\textwidth}
        \includegraphics[width=\textwidth]{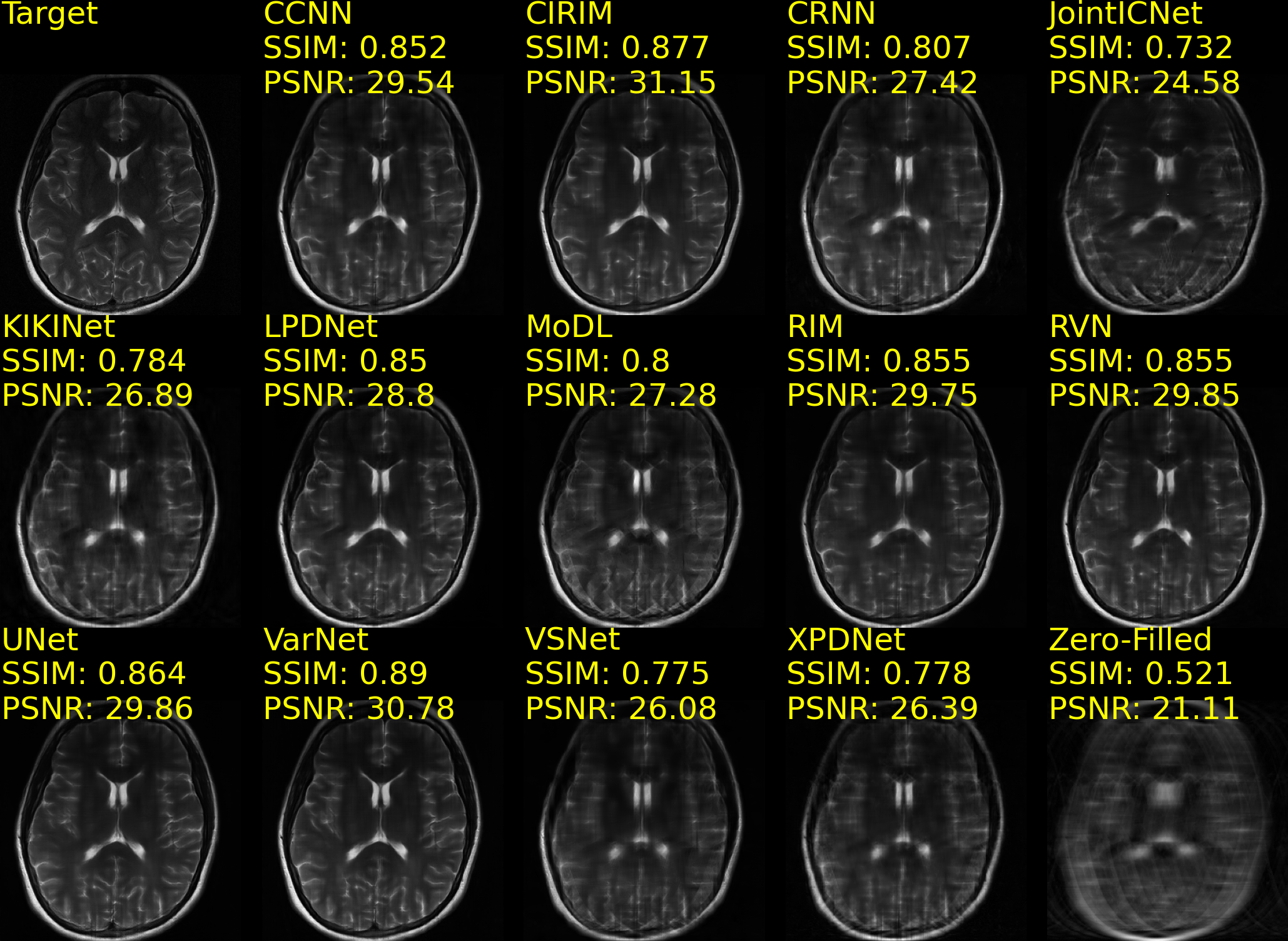}
        \caption{fastMRIBrains 14-coil data - 8x acceleration}
        \label{subfig:recon_fastmri_brains-8x}
    \end{subfigure}
    \caption{Reconstructions of 14-coil T$_2$-weighted data from the fastMRI Brains dataset, undersampled with an Equispaced 1D sampling pattern for 4x (Fig. \ref{subfig:recon_fastmri_brains-4x}) and 8x (Fig. \ref{subfig:recon_fastmri_brains-8x}) acceleration. The top row-first column shows the ground truth (Target) image. SSIM and PSNR scores are reported for each method and computed against the Target image. Methods are sorted alphabetically.}
    \label{fig:recon_fastmri_brains}
\end{figure}

\begin{figure}[!ht]
    \centering
    \includegraphics[width=0.485\textwidth]{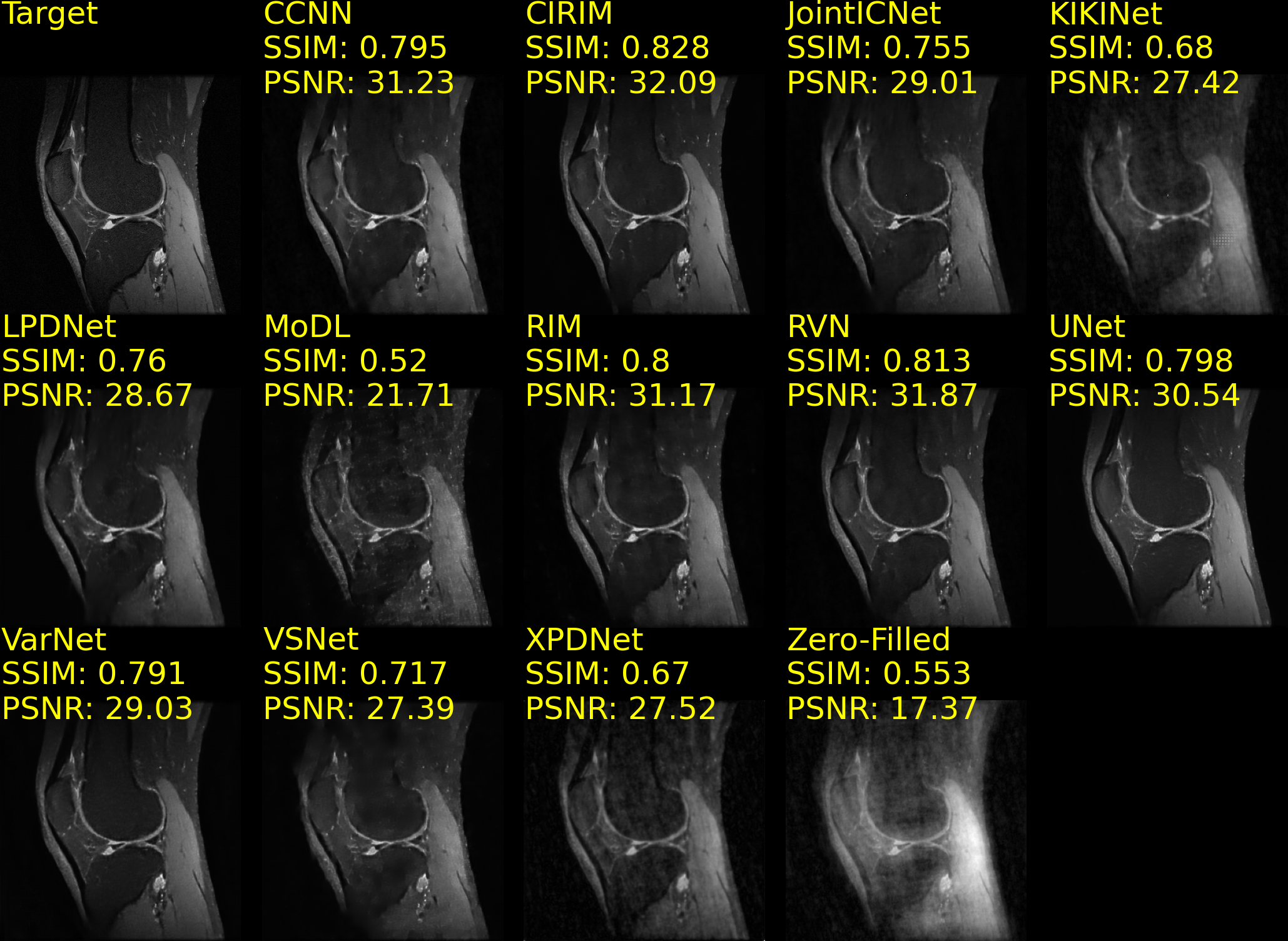}
    \caption{Reconstructions of 8-coil T$_2$-weighted Fast-Spin Echo data from the Stanford Knee dataset undersampled with a Gaussian 2D sampling pattern for 12x acceleration. The top row-first column shows the ground truth (Target) image. SSIM and PSNR scores are reported for each method and computed against the Target image. Methods are sorted alphabetically.}
    \label{fig:recon_stanford_knees}
\end{figure}

Example reconstructions of brain data are shown in Fig. \ref{subfig:recon_cc359-5x} and Fig. \ref{subfig:recon_cc359-10x}, from the CC359 dataset, and Fig. \ref{subfig:recon_fastmri_brains-4x} and Fig. \ref{subfig:recon_fastmri_brains-8x}, from the fastMRIBrain dataset. Figure \ref{fig:recon_stanford_knees} shows example reconstructions of knee data from the StanfordKnee dataset.

\begin{figure}[!ht]
    \centering
    \includegraphics[width=0.4\textwidth]{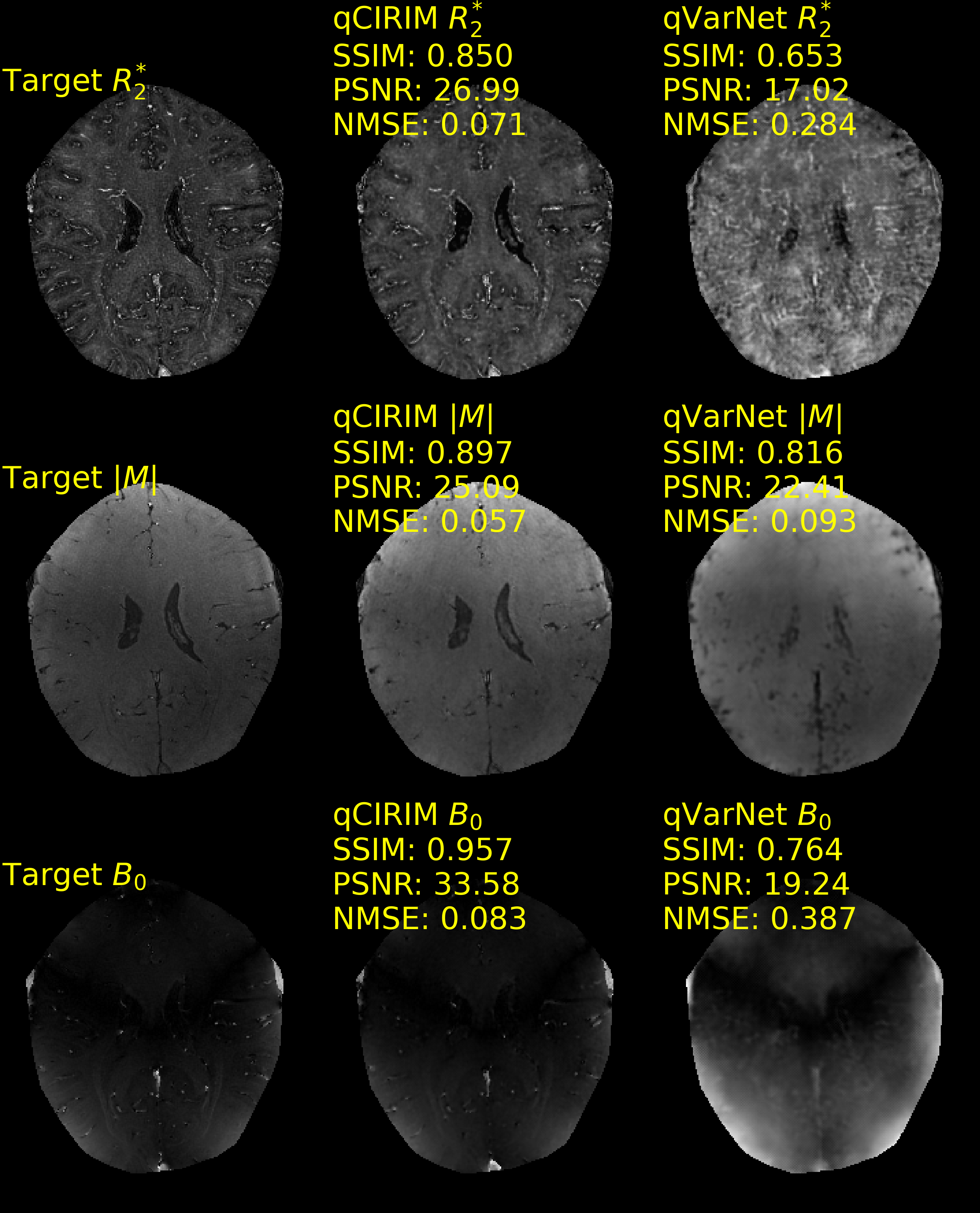}
    \caption{Quantitative parameter map estimation of 32-coil T$_1$-weighted data from the AHEAD dataset undersampled with a Gaussian 2D sampling pattern for 12x acceleration. The first column shows the ground truth (Target) quantitative parameter maps, $R^*_{2}$, $|M|$, and $B_0$ from top to bottom, respectively. The CIRIM and the VarNet were first used to reconstruct the undersampled data and give them as inputs to the qCIRIM and the qVarNet, respectively, to estimate the quantitative parameter maps, as shown in the second and third columns. SSIM, PSNR, and NMSE scores are reported for each method and computed against the Target quantitative parameter map.}
    \label{fig:ahead-qmri}
\end{figure}

Table \ref{tab:qmri} reports the performance of models trained on the AHEAD dataset for reconstruction and quantitative parameter map estimation. The CIRIM scored highest on reconstructing the AHEAD data for 12x acceleration, resulting in better initializations for the quantitative CIRIM (qCIRIM) model and, thus, more accurate quantitative parameter map estimation than the VarNet. The qCIRIM outperformed the quantitative VarNet (qVarNet) on accurately approximating the $R^*_{2}$, $B_0$, and $|M|$ quantitative maps. Example quantitative map estimations are shown in Fig. \ref{fig:ahead-qmri}.

\begin{table}[!ht]
    \caption{Overview of performance on reconstructing and estimating quantitative parameter maps. The AHEAD dataset was used, while data were 12x accelerated with a Gaussian 2D undersampling pattern. In the first column, the name of the model is reported. Each model's SSIM, PSNR, and NMSE scores are reported in the second, third, and fourth columns. Up arrows indicate the highest, the best, and down arrows indicate the lowest, the best. The performance of the reconstruction models is reported in the fourth and fifth row, while the quantitative parameter map estimation models' performance is reported in the seventh and eighth row. Best performing models are highlighted in bold. Methods are sorted in alphabetical order.}
    \label{tab:qmri}
    \centering
    \begin{tabularx}{\textwidth}{l*{3}{>{\centering\arraybackslash}X}}
        Model & \multicolumn{3}{c}{AHEAD - Gaussian 2D - 12x} \\
         & SSIM $\uparrow$ & PSNR $\uparrow$ & NMSE $\downarrow$ \\
        \midrule
         & \multicolumn{3}{c}{Reconstruction} \\
        \midrule
        \textbf{CIRIM} & \textbf{0.910 $\pm$ 0.077} & \textbf{32.86 $\pm$ 8.51} & \textbf{0.043 $\pm$ 0.065} \\
        VarNet     & 0.893     $\pm$ 0.055  & 32.37     $\pm$ 4.88  & 0.047     $\pm$ 0.054 \\
        \midrule
        & \multicolumn{3}{c}{Quantitative parameter map estimation} \\
        \midrule
        \textbf{qCIRIM} & \textbf{0.881 $\pm$ 0.178} & \textbf{28.36 $\pm$ 11.55} & \textbf{0.124 $\pm$ 0.363} \\
        qVarNet     & 0.784     $\pm$ 0.206  & 24.35     $\pm$ 7.77   & 0.192     $\pm$ 0.334 \\
    \end{tabularx}
\end{table}

\begin{figure}[!ht]
    \centering
    \begin{subfigure}[t]{\textwidth}
        \includegraphics[width=\textwidth]{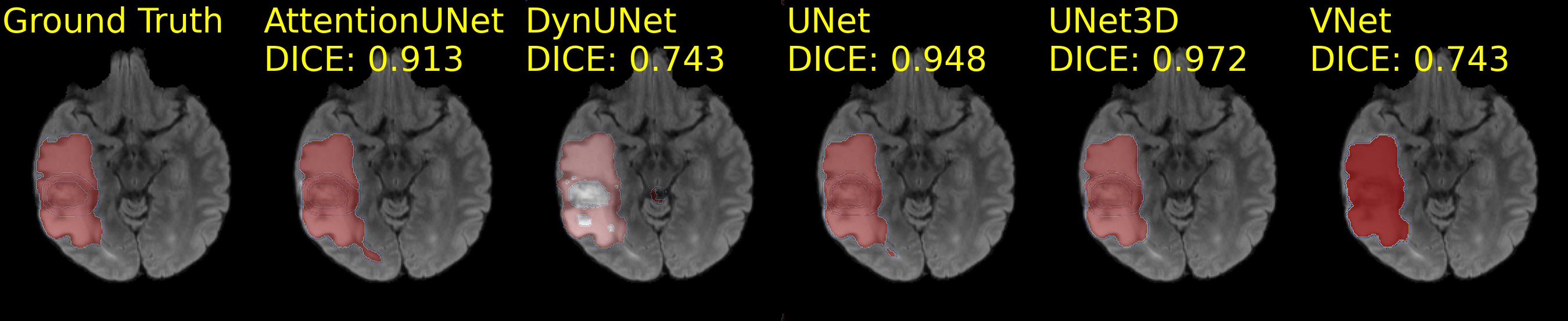}
        \caption{Brain Tumor Segmentation 2023 Adult Glioma segmentations.}
        \label{subfig:brats}
    \end{subfigure}
    \begin{subfigure}[t]{\textwidth}
        \includegraphics[width=\textwidth]{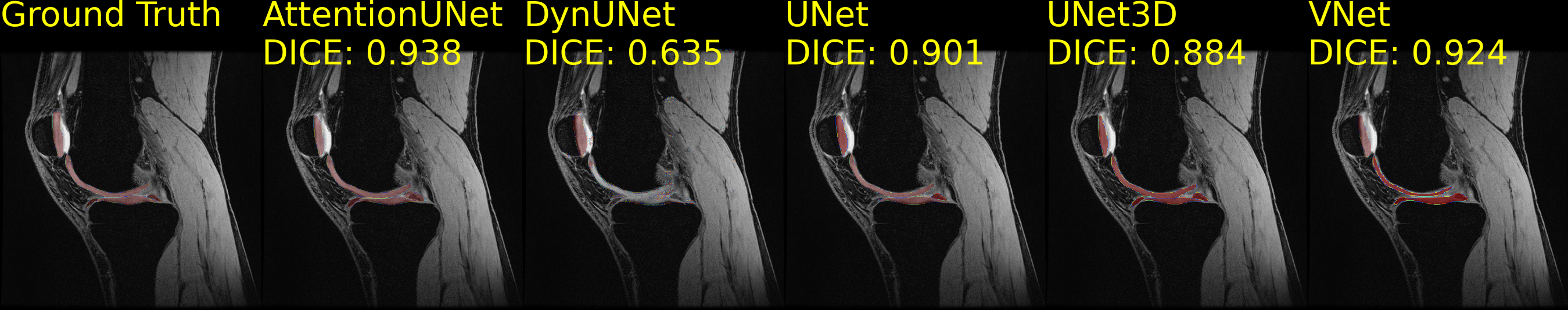}
        \caption{Stanford Knee MRI with Multi-Task Evaluation segmentations.}
        \label{subfig:SKM-TEA-seg}
    \end{subfigure}
    \begin{subfigure}[t]{\textwidth}
        \includegraphics[width=\textwidth]{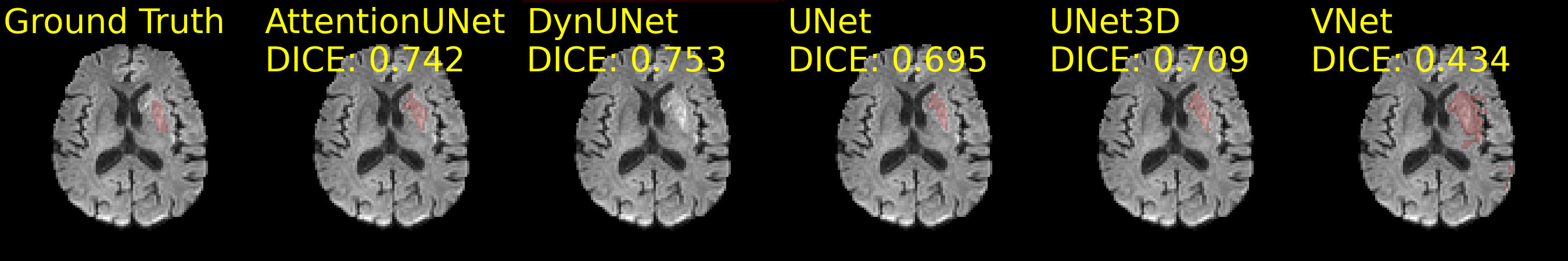}
        \caption{ISLES 2022 Sub Acute Stroke segmentations.}
        \label{subfig:isles}
    \end{subfigure}
    \caption{Segmentations on the Brain Tumor Segmentation 2023 Adult Glioma dataset (Fig. \ref{subfig:brats}), the Stanford Knee MRI with Multi-Task Evaluation (SKM-TEA) segmentation-only dataset (Fig. \ref{subfig:SKM-TEA-seg}), and the ISLES 2022 Sub Acute Stroke dataset (Fig. \ref{subfig:isles}). The first image on each figure shows the Ground Truth image with the segmentation labels. The rest of the images present the segmentations of different methods. Each segmentation method's DICE score is reported and computed against the Ground Truth labels. Methods are sorted alphabetically.}
    \label{fig:segs}
\end{figure}

Table \ref{tab:brats_SKM-TEA} presents the performance in the segmentation task of models trained on the BraTS2023AdultGlioma and the SKM-TEA segmentation-only datasets. On the BraTS2023AdultGlioma dataset, the UNet achieved the highest DICE score and the lowest HD95 score, while the UNet3D achieved the highest F1 score and the AttentionUNet the highest IOU. Although the high DICE scores, F1 and IOU scores were reported lower, which may be attributed to the heterogeneous tumors, leading to the inclusion of non-tumor regions in the predicted segmentation. Similar observations were made on the SKM-TEA segmentation-only dataset, where the UNet3D and VNet achieved the highest DICE score. While the UNet3D scored the highest IOU, the VNet scored the lowest HD95 and the highest F1. In general, the lower F1 scores may be attributed to the heterogeneity of the data since they were acquired from multiple vendors (Sec. \ref{sec:datasets}). The variability of knee structures across different patients could cause low IOU scores. The impact on the variability between high DICE scores and low F1 and IOU scores can be seen in Fig. \ref{subfig:brats} and Fig. \ref{subfig:SKM-TEA-seg}, where the DynUNet underestimated the segmented classes, resulting on the worst-performing model. 

Table \ref{tab:isles} reports the performance of segmentation models on the ISLES2022SubAcuteStroke dataset. The DynUNet achieved the lowest average lesion distance (ALD) and the highest DICE and Lesion-F1 scores. The UNet achieved the lowest average volume difference (AVD). The worst-performing model was the VNet, which overestimated the lesion segmentation, as shown in Fig. \ref{subfig:isles}.

\begin{table}[!ht]
    \caption{Overview of performance in the segmentation task of models trained on the BraTS2023AdultGlioma (third to seventh row) and the SKM-TEA segmentation-only (ninth to thirteenth row) datasets. Model name, DICE, F1, Hausdorff Distance 95\% (HD95), and Intersection Over Union (IOU) scores are reported from left to right. Up and down arrows indicate higher and lower scores being better, respectively. Best performing models are highlighted in bold. Methods are sorted in alphabetical order.}
    \label{tab:brats_SKM-TEA}
    \centering
    \begin{tabular}{ccccc}
        \toprule
        Model & DICE $\uparrow$ & F1 $\uparrow$ & HD95 $\downarrow$ & IOU $\uparrow$ \\
        \midrule
        & \multicolumn{3}{c}{BraTS 2023 Adult Glioma} & \\
        \midrule
        \textbf{AttentionUNet} & 0.930     $\pm$ 0.126  & 0.648     $\pm$ 0.763  & 3.836     $\pm$ 3.010  & \textbf{0.537 $\pm$ 0.662} \\ 
        DynUNet            & 0.806     $\pm$ 0.276  & 0.104     $\pm$ 0.580  & 5.119     $\pm$ 5.411  & 0.070     $\pm$ 0.419 \\ 
        \textbf{UNet}          & \textbf{0.937 $\pm$ 0.118} & 0.671     $\pm$ 0.787  & \textbf{3.504 $\pm$ 2.089} & 0.535     $\pm$ 0.663 \\ 
        \textbf{UNet3D}        & 0.936     $\pm$ 0.133  & \textbf{0.674 $\pm$ 0.782} & 3.550     $\pm$ 2.162  & 0.528     $\pm$ 0.652 \\ 
        VNet               & 0.733     $\pm$ 0.437  & 0.014     $\pm$ 0.234  & 6.010     $\pm$ 6.097  & 0.000     $\pm$ 0.004 \\ 
        \midrule
        & \multicolumn{3}{c}{SKM-TEA segmentation-only} & \\
        \midrule
        AttentionUNet & 0.909     $\pm$ 0.088  & 0.637     $\pm$ 0.475  & 6.358     $\pm$ 2.209  & 0.529     $\pm$ 0.361 \\ 
        DynUNet       & 0.689     $\pm$ 0.136  & 0.059     $\pm$ 0.264  & 8.973     $\pm$ 4.507  & 0.015     $\pm$ 0.066 \\
        UNet          & 0.912     $\pm$ 0.058  & 0.651     $\pm$ 0.449  & 6.618     $\pm$ 1.793  & 0.516     $\pm$ 0.350 \\ 
        \textbf{UNet3D}   & \textbf{0.918 $\pm$ 0.068} & 0.789     $\pm$ 0.404  & 5.893     $\pm$ 2.995  & \textbf{0.530 $\pm$ 0.347} \\ 
        \textbf{VNet}     & \textbf{0.918 $\pm$ 0.081} & \textbf{0.816 $\pm$ 0.426} & \textbf{5.540 $\pm$ 3.036} & 0.507     $\pm$ 0.388 \\ 
        \bottomrule
    \end{tabular}
\end{table}

\begin{table}[!ht]
    \caption{Overview of performance in the segmentation task of models trained on the ISLES2022SubAcuteStroke dataset. Model name, Absolute Lesion Difference (AVD), Absolute Volume Difference (AVD), DICE, and Lesion F1 (L-F1) scores are reported from left to right. Up and down arrows indicate whether higher or lower values indicate better performance. Best performing models are highlighted in bold. Methods are sorted in alphabetical order.}
    \centering
    \begin{tabular}{ccccc}
        \toprule
        Model & \multicolumn{3}{c}{ISLES 2022 Sub Acute Stroke} \\
         & ALD $\downarrow$ & AVD $\downarrow$ & DICE $\uparrow$ & L-F1 $\uparrow$ \\
        \midrule
        AttentionUNet & 0.809     $\pm$ 2.407      & 0.548     $\pm$ 3.411       & 0.709     $\pm$ 0.552      & 0.799     $\pm$ 0.579 \\ 
        \bf{DynUNet}  & \bf{0.752 $\pm$ 2.230}     & 0.586     $\pm$ 3.874       & \bf{0.729 $\pm$ 0.529}     & \bf{0.802 $\pm$ 0.564} \\ 
        \bf{UNet}     & 0.909     $\pm$ 3.953      & \bf{0.544 $\pm$ 3.921}      & 0.695     $\pm$ 0.559      & 0.786     $\pm$ 0.585 \\ 
        UNet3D        & 0.821     $\pm$ 2.167      & 0.691     $\pm$ 5.458       & 0.687     $\pm$ 0.547      & 0.798     $\pm$ 0.573 \\ 
        VNet          & 2.281     $\pm$ 10.72      & 3.257     $\pm$ 27.430      & 0.490     $\pm$ 0.694      & 0.600     $\pm$ 0.687 \\ 
    \end{tabular}
    \label{tab:isles}
\end{table}

\begin{figure}[!ht]
    \centering
    \includegraphics[width=\textwidth]{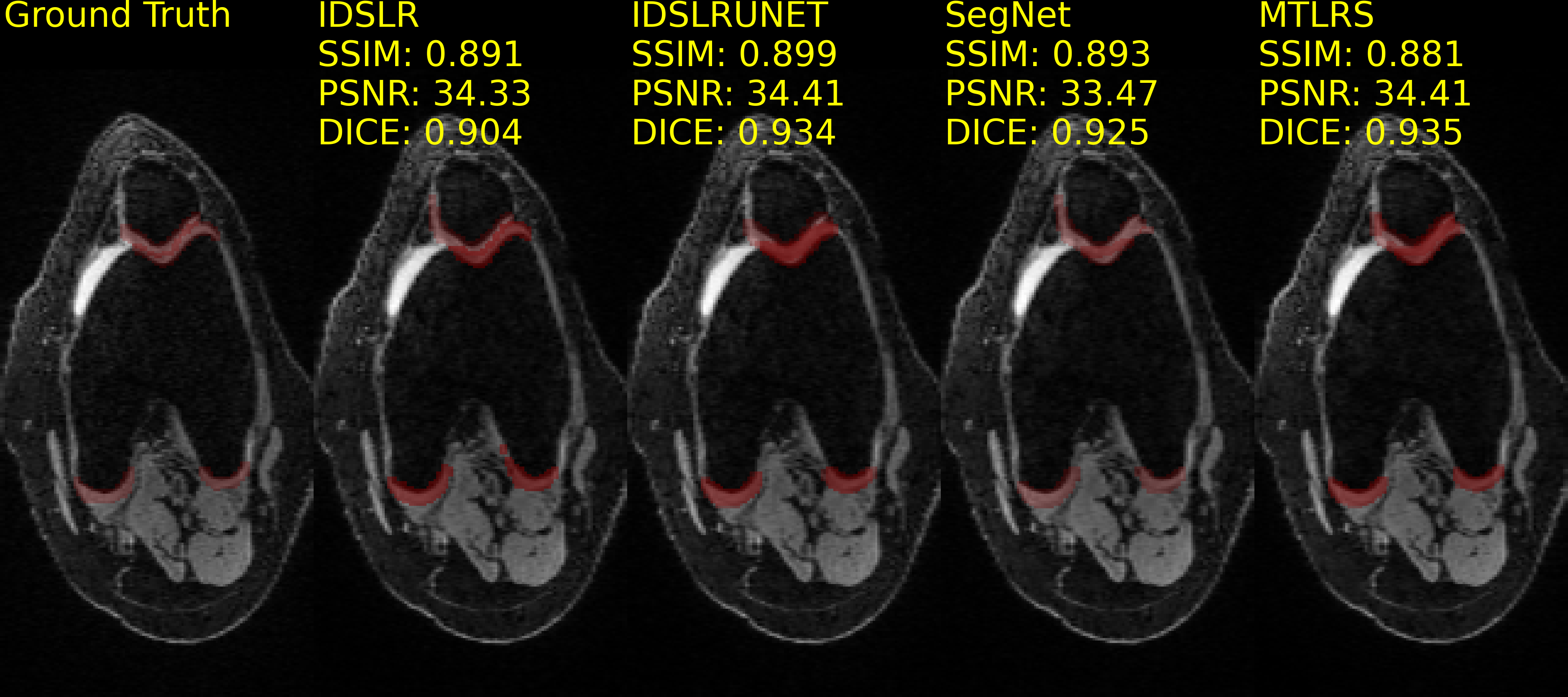}
    \caption{Reconstructions and segmentations on the Stanford Knee MRI with Multi-Task Evaluation (SKM-TEA) dataset, undersampled with a Poisson disc distribution 2D pattern for 4x acceleration. First image shows the Ground Truth image and segmentation labels. SSIM, PSNR, and DICE scores are reported for each method and computed against the Ground Truth image and segmentation labels. Methods are sorted alphabetically. Images are interpolated for visualisation purposes from $416 \times 80$ to $256 \times 128$.}
    \label{fig:SKM-TEA-seg-mtl}
\end{figure}

Table \ref{tab:mtlrs} reports the performance in MTL for joint reconstruction and segmentation models trained on the SKM-TEA dataset. The IDSLRUNET achieved the highest SSIM and PSNR scores, while the SegNet achieved the highest DICE, F1, and IOU and lowest HD95 scores. The lower F1 and IOU scores, as also observed in the performance of the segmentation-only models, may be attributed to varying patient anatomy, data acquisition, and different knee structures. Example reconstructions and segmentations when performing MTL can be found in Fig. \ref{fig:SKM-TEA-seg-mtl}.

\begin{table*}[!ht]
    \caption{Overview of performance in MTL for joint reconstruction and segmentation of models trained on the SKM-TEA dataset for Poisson 2D 4x undersampling. Model name, SSIM, PSNR, DICE, F1, Hausdorff Distance 95\% (HD95), and Intersection Over Union (IOU) scores are reported from left to right. Up and down arrows indicate whether higher or lower scores indicate higher performance. Best performing models are highlighted in bold. Methods are sorted in alphabetical order.}
    \centering
    \begin{adjustbox}{width=\textwidth}
        \begin{tabular}{cccccccc}
            Model & \multicolumn{5}{c}{SKM-TEA - Poisson 2D 4x} \\
             & SSIM $\uparrow$ & PSNR $\uparrow$ & DICE $\uparrow$ & F1 $\uparrow$ & HD95 $\downarrow$ & IOU $\uparrow$ \\
            \midrule
            IDSLR          & 0.836     $\pm$ 0.106  & 30.38     $\pm$ 5.67  & 0.894     $\pm$ 0.127  & 0.256     $\pm$ 0.221  & 4.927     $\pm$ 2.812  & 0.298     $\pm$ 0.309 \\ 
            \bf{IDSLRUNET} & \bf{0.842 $\pm$ 0.106} & \bf{30.53 $\pm$ 5.59} & 0.870     $\pm$ 0.134  & 0.225     $\pm$ 0.194  & 8.724     $\pm$ 3.298  & 0.212     $\pm$ 0.199 \\ 
            MTLRS          & 0.832     $\pm$ 0.106  & 30.48     $\pm$ 5.30  & 0.889     $\pm$ 0.118  & 0.247     $\pm$ 0.203  & 7.594     $\pm$ 3.673  & 0.218     $\pm$ 0.194 \\ 
            \bf{SegNet}    & 0.840     $\pm$ 0.107  & 29.95     $\pm$ 5.12  & \bf{0.915 $\pm$ 0.114} & \bf{0.270 $\pm$ 0.284} & \bf{3.002 $\pm$ 1.449} & \bf{0.290 $\pm$ 0.349} \\ 
        \end{tabular}
    \end{adjustbox}
    \label{tab:mtlrs}
\end{table*}

\section{Discussion and Conclusions} 
\label{sec:discussion}

We presented the Advanced Toolbox for Multitask Medical Imaging Consistency (ATOMMIC), a versatile toolbox designed to ensure consistency in the performance of various Deep Learning (DL) models applied in different MRI tasks such as reconstruction, quantitative parameter map estimation and segmentation (Sec. \ref{sec:mri-tasks}). Consistency is ensured by unifying the implementation of networks' components, hyperparameters, image transformations, and training configurations. Among existing AI frameworks for MRI analysis, ATOMMIC was found to be the only framework to uniquely harmonize complex-valued and real-valued data support, allowing the assessment of MultiTask Learning (MTL) by combining individual models designed for single tasks to perform joint tasks. To demonstrate ATOMMIC's capabilities we trained and tested twenty-five DL models on eight publicly available datasets, including brain and knee anatomies, for three different MRI tasks and presented applications of MTL for joint reconstruction and segmentation. Three undersampling schemes were evaluated, ranging from 4 times to 12 times acceleration, on the task of accelerated MRI reconstruction on three publicly available datasets, using different loss functions, optimizers, and learning rate schedulers (Fig \ref{fig:atommic_schematic_overview}). We also assessed the effectiveness of end-to-end reconstruction and coil sensitivity maps estimation during training, removing the overhead of pre-computing coil sensitivity maps and increasing the storage space. Coil compression was evaluated on the reconstruction task, showing advantages in reducing training time while maintaining high performance (Sec. \ref{sec:mri-transforms}). Successful application of quantitative DL models was demonstrated in accurately estimating quantitative parameter maps of the brain, such as the $R^*_{2}$ map, which allows for quantifying iron deposition related to aging and Parkinson's and Alzheimer's disease (\cite{zhangUnifiedModelReconstruction2022}). Segmentation of brain lesions, tumors, and knee pathologies was also presented, while the tasks of segmentation and reconstruction were combined to assess MTL.

Physics-based DL models outperformed other DL models on the task of accelerated MRI reconstruction, showing an advantage in enforcing data consistency on the MRI domain either implicity (CIRIM) or explicitly (JointICNet, RVN, VarNet). The models trained and tested on the fastMRIBrains (Table \ref{tab:recon_cc359_fastMRIBrains}) and StanfordKnee datasets (Table \ref{tab:recon_stanford_knees}) exhibited higher standard deviations than the models trained on the CC359 dataset (Table \ref{tab:recon_cc359_fastMRIBrains}), potentially due to the different undersampling patterns, multiple modalities, and varying numbers of coils for the fastMRIBrains dataset and the small number of coils (eight) for the StanfordKnee dataset (Sec. \ref{sec:datasets}). These variations in data acquisition resulted in decreased SNR, adversely impacting the SSIM and PSNR scores. Physics-based models were also demonstrated to accurately approximate quantitative parameter maps (Table \ref{tab:qmri}). This suggests that maintaining consistency, an essential feature in ATOMMIC, in the primary task of reconstruction can enhance performance in the subsequent quantitative parameter map estimation task, a promising advancement towards fast and robust quantification of neurological diseases. The baseline UNet performed the best on the BraTS2023AdultGlioma and SKM-TEA datasets on the segmentation task (Table \ref{tab:brats_SKM-TEA}), but F1 and IOU scores were low for all models, potentially due to the datasets' heterogeneity and distribution shifts. The UNet3D and the VNet showed strengths in handling heterogeneous data. However, the VNet struggled with small lesion segmentation in the ISLES2022SubAcuteStroke dataset (Table \ref{tab:isles}). The AttentionUNet achieved the highest IOU score on the BraTS2023AdultGlioma dataset, underscoring its ability to identify relevant tumor regions, although heterogeneous due to the integrated attention mechanisms. The dynamic nature of the DynUNet proved advantageous in the ISLES2022SubAcuteStroke dataset, where it outperformed others, scoring the highest ALD, DICE, and Lesion-F1 scores. ATOMMIC's advancement in utilizing MTL for combining tasks led to improved reconstruction quality and segmentation accuracy when performing the two tasks jointly (Table \ref{tab:mtlrs}).

Evaluating multiple DL models on a single task using a robust framework provides a better understanding of the benefits of applying DL to medical imaging rather than just focusing on the performance of a single model. Utilizing MTL to combine tasks is a step towards end-to-end solutions that eliminate the overhead of splitting related tasks, leading to improved performance and faster processing speed. ATOMMIC enables the evaluation of many DL models on multiple public datasets with standardized formats, thanks to the significant efforts made by various research groups. However, supporting private datasets can be challenging. In addition to privacy concerns, identifying a series of appropriate pre-processing steps to use the data is often necessary, which can be time-consuming and require expert knowledge. Raw MRI data, in particular, often come in vendor-locked proprietary formats. While the ISMRM-RD format (\cite{inatiISMRMRawData2017}) represents a step towards an open vendor-agnostic format for storing such data, integrating private datasets into open-source toolboxes remains limited. Such limitation is also identified in our work, and further limitations include the fact that ATOMMIC currently focuses solely on MRI, and essential tasks such as classification, registration, and motion correction remain to be implemented towards a robust end-to-end multitask framework.

Future work could focus on developing pre-processing pipelines for private medical imaging datasets, including additional tasks and imaging modalities, such as Computed Tomography. Nevertheless, open issues like data interoperability, model robustness, and assessing MTL still require attention. The availability of public datasets, open-source code, and comprehensive documentation is crucial to effectively adopting AI techniques and facilitating their further development by the scientific and broader research communities. With ATOMMIC, we aim to accelerate medical image analysis and provide a comprehensive framework for researchers to integrate and evaluate datasets, DL models, tasks, and imaging modalities.

\section*{Acknowledgments}
This publication is based on the STAIRS project under the TKI-PPP program. The collaboration project is co-funded by the PPP Allowance made available by Health~Holland, Top Sector Life Sciences \& Health, to stimulate public-private partnerships.

H.A. Marquering and M.W.A. Caan are shareholders of Nicolab International Ltd. H.A. Marquering is a shareholder of TrianecT B.V. and inSteps B.V. (unrelated to this project; all paid individually).

%%Harvard
\bibliographystyle{unsrtnat}
% \bibliography{references}

\vfill \newpage

\section*{Appendix}

\begin{longtable}{cc}
    \caption{Overview of selected hyperparameters for all trained models on all tasks.} \\
    \toprule
    \textbf{Model} & \textbf{Hyperparameters} \\ \midrule \\
    \multicolumn{2}{c}{\textbf{Accelerated MRI Reconstruction}} \\ \\
    \bottomrule

    CCNN & 
    \begin{minipage}[t]{0.8\linewidth}
        num\_cascades: 10, hidden\_channels: 64, n\_convs: 5
    \end{minipage} \\ \midrule
    
    CIRIM & 
    \begin{minipage}[t]{0.8\linewidth}
        recurrent\_layer: IndRNN, conv\_filters: [128, 128, 2], conv\_kernels: [5, 3, 3], conv\_dilations: [1, 2, 1], conv\_bias: [true, true, false], recurrent\_filters: [128, 128, 0], recurrent\_kernels: [1, 1, 0], recurrent\_dilations: [1, 1, 0], recurrent\_bias: [true, true, false], time\_steps: 8, conv\_dim: 2, num\_cascades: 5
    \end{minipage} \\ \midrule
    
    CRNN & 
    \begin{minipage}[t]{0.8\linewidth}
        num\_iterations: 10, hidden\_channels: 64, n\_convs: 3
    \end{minipage} \\ \midrule
    
    JointICNet & 
    \begin{minipage}[t]{0.8\linewidth}
        num\_iter: 2, kspace\_unet\_num\_filters: 16, kspace\_unet\_num\_pool\_layers: 2, imspace\_unet\_num\_filters: 16, imspace\_unet\_num\_pool\_layers: 2, sens\_unet\_num\_filters: 16, sens\_unet\_num\_pool\_layers: 2
    \end{minipage} \\ \midrule
    
    KIKINet & 
    \begin{minipage}[t]{0.8\linewidth}
        num\_iter: 2, 
        kspace\_unet\_num\_filters: 16, 
        kspace\_unet\_num\_pool\_layers: 2, 
        imspace\_unet\_num\_filters: 16, 
        imspace\_unet\_num\_pool\_layers: 2
    \end{minipage} \\ \midrule
    
    LPDNet & 
    \begin{minipage}[t]{0.8\linewidth}
        num\_primal: 5, 
        num\_dual: 5, 
        num\_iter: 5, 
        primal\_unet\_num\_filters: 16, 
        primal\_unet\_num\_pool\_layers: 2, 
        dual\_unet\_num\_filters: 16
    \end{minipage} \\ \midrule
    
    MoDL & 
    \begin{minipage}[t]{0.8\linewidth}
        unrolled\_iterations: 5, 
        residual\_blocks: 5, 
        channels: 64, 
        regularization\_factor: 0.1
    \end{minipage} \\ \midrule
    
    RIM & 
    \begin{minipage}[t]{0.8\linewidth}
        recurrent\_layer: GRU, 
        conv\_filters: [64, 64, 2], 
        conv\_kernels: [5, 3, 3], 
        conv\_dilations: [1, 2, 1], 
        conv\_bias: [true, true, false], 
        recurrent\_filters: [64, 64, 0], 
        recurrent\_kernels: [1, 1, 0], 
        recurrent\_dilations: [1, 1, 0], 
        recurrent\_bias: [true, true, false], 
        time\_steps: 8, 
        conv\_dim: 2
    \end{minipage} \\ \midrule
    
    RVN & 
    \begin{minipage}[t]{0.8\linewidth}
        recurrent\_hidden\_channels: 64, 
        recurrent\_num\_layers: 4, 
        num\_steps: 8, 
        learned\_initializer: true, 
        initializer\_initialization: "sense", 
        initializer\_channels: [32, 32, 64, 64], 
        initializer\_dilations: [1, 1, 2, 4]
    \end{minipage} \\ \midrule
    
    UNet & 
    \begin{minipage}[t]{0.8\linewidth}
        channels: 64, 
        pooling\_layers: 4
    \end{minipage} \\ \midrule
    
    VarNet & 
    \begin{minipage}[t]{0.8\linewidth}
        num\_cascades: 8, 
        channels: 18, 
        pooling\_layers: 4
    \end{minipage} \\ \midrule
    
    VSNet & 
    \begin{minipage}[t]{0.8\linewidth}
        num\_cascades: 10, 
        imspace\_model\_architecture: CONV, 
        imspace\_conv\_hidden\_channels: 64, 
        imspace\_conv\_n\_convs: 4
    \end{minipage} \\ \midrule
    
    XPDNet & 
    \begin{minipage}[t]{0.8\linewidth}
        num\_primal: 5, 
        num\_dual: 1, 
        num\_iter: 10, 
        use\_primal\_only: true, 
        kspace\_model\_architecture: CONV, 
        image\_model\_architecture: MWCNN, 
        mwcnn\_hidden\_channels: 16, 
        mwcnn\_bias: true
    \end{minipage} \\

    \bottomrule \\ 
    
    \multicolumn{2}{c}{\textbf{quantitative MRI parameter map estimation}} \\ \\
    \bottomrule

    qCIRIM & 
    \begin{minipage}[t]{0.8\linewidth}
        quantitative\_module\_recurrent\_layer: IndRNN, quantitative\_module\_conv\_filters: [128, 128, 2], quantitative\_module\_conv\_kernels: [5, 3, 3], quantitative\_module\_conv\_dilations: [1, 2, 1], quantitative\_module\_conv\_bias: [true, true, false], quantitative\_module\_recurrent\_filters: [128, 128, 0], quantitative\_module\_recurrent\_kernels: [1, 1, 0], quantitative\_module\_recurrent\_dilations: [1, 1, 0], quantitative\_module\_recurrent\_bias: [true, true, false], quantitative\_module\_time\_steps: 8, quantitative\_module\_conv\_dim: 2, quantitative\_module\_num\_cascades: 5
    \end{minipage} \\ \midrule
    
    qVarNet & 
    \begin{minipage}[t]{0.8\linewidth}
        quantitative\_module\_num\_cascades: 8, 
        quantitative\_module\_channels: 18, 
        quantitative\_module\_pooling\_layers: 4
    \end{minipage} \\

    \bottomrule
    \\
    
    \multicolumn{2}{c}{\textbf{MRI Segmentation}} \\ \\
    \bottomrule

    AttentionUNet & 
    \begin{minipage}[t]{0.8\linewidth}
        channels: 32, 
        pooling\_layers: 5
    \end{minipage} \\ \midrule

    DynUNet & 
    \begin{minipage}[t]{0.8\linewidth}
        channels: 32, 
        pooling\_layers: 5, 
        activation: leakyrelu, 
        deep\_supervision: true, 
        deep\_supervision\_levels: 2
    \end{minipage} \\ \midrule

    UNet & 
    \begin{minipage}[t]{0.8\linewidth}
        channels: 32, 
        pooling\_layers: 5
    \end{minipage} \\ \midrule

    UNet3D & 
    \begin{minipage}[t]{0.8\linewidth}
        channels: 32, 
        pooling\_layers: 5, 
        consecutive\_slices: 3
    \end{minipage} \\ \midrule

    VNet & 
    \begin{minipage}[t]{0.8\linewidth}
        channels: 16, 
        pooling\_layers: 5, 
        activation: elu
    \end{minipage} \\

    \bottomrule
    \\

    \multicolumn{2}{c}{\textbf{MultiTask Learning for joint Accelerated MRI Reconstruction \& MRI Segmentation}} \\ \\
    \bottomrule

    IDSLR & 
    \begin{minipage}[t]{0.8\linewidth}
        channels: 64, 
        pooling\_layers: 2, 
        num\_iters: 5
    \end{minipage} \\ \midrule

    IDSLRUNET & 
    \begin{minipage}[t]{0.8\linewidth}
        channels: 64, 
        pooling\_layers: 2, 
        num\_iters: 5
    \end{minipage} \\ \midrule

    MTLRS & 
    \begin{minipage}[t]{0.8\linewidth}
        num\_cascades: 5, 
        reconstruction\_module\_recurrent\_layer: IndRNN, 
        reconstruction\_module\_conv\_filters: [64, 64, 2], 
        reconstruction\_module\_conv\_kernels: [5, 3, 3], 
        reconstruction\_module\_conv\_dilations: [1, 2, 1], 
        reconstruction\_module\_conv\_bias: [true, true, false], 
        reconstruction\_module\_recurrent\_filters: [64, 64, 0], 
        reconstruction\_module\_recurrent\_kernels: [1, 1, 0], 
        reconstruction\_module\_recurrent\_dilations: [1, 1, 0], 
        reconstruction\_module\_recurrent\_bias: [true, true, false], 
        reconstruction\_module\_time\_steps: 8, 
        reconstruction\_module\_conv\_dim: 2, 
        reconstruction\_module\_num\_cascades: 1, 
        segmentation\_module: AttentionUNet, 
        segmentation\_module\_channels: 32, 
        segmentation\_module\_pooling\_layers: 5
    \end{minipage} \\ \midrule

    SEGNET & 
    \begin{minipage}[t]{0.8\linewidth}
        channels: 64, 
        pooling\_layers: 2, 
        num\_cascades: 5, 
        final\_layer\_conv\_dim: 2, 
        final\_layer\_kernel\_size: 3, 
        final\_layer\_dilation: 1, 
        final\_layer\_bias: False, 
        final\_layer\_nonlinear: relu 
    \end{minipage} \\
    
    \bottomrule
\label{tab:hp-overview}
\end{longtable}

\end{document}